\documentclass[acmsmall,screen,nonacm]{acmart}

\usepackage{mathpartir}
\usepackage{booktabs}

\DeclareUrlCommand\artifact{\urlstyle{tt}\small}

\newcommand{\ftrule}[1]{\textsc{#1}}
\newcommand{\calc}{\ensuremath{\Lambda_S}}
\newcommand{\Q}[1]{\ensuremath{\mathsf{Q}\,#1}}
\newcommand{\uall}[2]{\ensuremath{\forall #1{:}#2.\,}}
\newcommand{\dall}[1]{\ensuremath{\forall #1.\,}}
\newcommand{\cvt}[3]{\ensuremath{\mathsf{convert}\;#1\;#2\;#3}}
\newcommand{\ifle}[4]{\ensuremath{\mathsf{if}\;#1 \le #2\;\mathsf{then}\;#3\;\mathsf{else}\;#4}}
\newcommand{\dimOf}{\ensuremath{\mathsf{dim}}}

\usepackage{latexml}
\newcommand{\longcitep}[2]{\iflatexml(#1, \citeyearNP{#2})\else\citep{#2}\fi}
\iflatexml
  \renewenvironment{mathpar}{\let\and\relax}{}
  \renewcommand{\inferrule}[3][]{%
    \[\text{\textsc{#1}}\quad\dfrac{\let\\\qquad #2}{#3}\]}
  \renewcommand{\ifle}[4]{\ensuremath{\mathsf{if}\;#1 \le #2
    \mathbin{\mathsf{then}} #3 \mathbin{\mathsf{else}} #4}}
\fi

\begin{document}

\title{A Calculus for Units of Measure with Conversion}

\author{Eric Allen}
\email{eric.allen@twosigma.com}
\affiliation{\institution{Two Sigma Investments, LP}\city{New
York}\country{USA}}

\begin{abstract}
Programs that compute with physical quantities often need to convert
between units of measurement, and as the Mars Climate Orbiter showed,
these conversions can be a rich source of errors. Meanwhile, typed unit
calculi, our most rigorous formalisms for checking physical units in a
program, have excluded unit conversions. Through this exclusion, they
have established a powerful property: well-typed programs are invariant
under rescaling (no program can depend on how big a meter is). But
losing the ability to convert between units is a significant cost. In
contrast, practical languages provide conversion but no invariance
theorem. We present \calc, a typed lambda calculus with conversion,
quantification over units and dimensions, and vectors and linear maps
with per-component units, and we determine exactly how much invariance
survives conversion. For terms without unit constants we prove two
abstraction theorems: (\emph{i}) a convert-free term is invariant under
every rescaling; (\emph{ii}) a term with conversions is invariant under
every rescaling that scales all units of one dimension by the same
factor. The condition in (\emph{ii}) cannot be weakened: under any other
rescaling, some conversion of a nonzero value is not invariant.

For first-order programs, a verified decision procedure returns one of
three verdicts: it certifies that no error in the declared conversion
factors can change the program's answer, names the accumulated ratio
through which such an error would scale it, or declines. A verified
checker decides whether unit declarations are consistent and determine
every conversion factor, and extracts each factor exactly. We also prove
adequacy, erasure, and the $n$-variable Pi theorem of dimensional
analysis with conversion. Every theorem is mechanized in Lean~4, and the
evaluator compiles to a native binary.
\end{abstract}

\keywords{units of measure, conversion, parametricity, coherence, dimensional
analysis, mechanized metatheory, Lean}

\maketitle

\section{Introduction}

Programs that manipulate physical quantities must eventually convert between
units, and conversion is where the errors occur. The loss of the Mars Climate
Orbiter was traced to ground software that reported thruster impulse in
pound-seconds where its interface specified
newton-seconds~\longcitep{Mars Climate Orbiter Mishap Investigation Board}{stephenson1999}; the enduring popularity of units-of-measure
extensions, libraries, and checkers testifies that the problem is neither
exotic nor solved~\citep[a preprint]{danish2024}. Nor is the disaster a relic: a
lightweight detector found physical-unit inconsistencies in 11\% of 213
open-source robotics systems~\citep{ore2017}, and a units checker applied to the Weather
Research and Forecasting model found an equation converting grams to
micrograms by multiplying by $10^{-6}$ instead of
$10^{6}$~\citep{danish2024}.

At least superficially, the remedy looks like a type system, and an elegant
theoretical literature has explored this approach. Moreover, the static analysis
required has compiler applications beyond just spotting errors.
To \emph{rescale} a unit is to
multiply every quantity stored at it by a factor. A
numerical kernel may choose to rescale measurements in order to keep the
values it works with inside the range where floating point carries full
precision, or move a whole computation into one system of units so that no
operation has to convert. Both are choices of internal representation,
invisible in the source, and both are rescalings.\footnote{Binary powers are
useful choices of factor. IEEE~754~\citep{ieee754-2019}
represents a binary number by a significand and an exponent, so rescaling
by $2^k$ is exact when the value remains in the normal range.}

An \emph{abstraction theorem} says that well-typed programs respect
rescalings: rescale a program's inputs by their units' factors and its output
rescales by exactly the factor its result type prescribes. No program can
depend on how big a meter is. Beginning with \citet{kennedy1997}, typed
calculi with units as a free abelian group enjoy abstraction theorems,
non-definability results (e.g., no term of type
$\forall \mathrm{u}.\,\Q{\mathrm{u}^2} \to \Q{\mathrm{u}}$ computes square root), and the Pi theorem of
dimensional analysis, all as consequences of parametricity. The Pi theorem
says that a dimensionally consistent relation
among $n$ quantities factors through their dimensionless
combinations. But
that literature rests on a design property: \emph{nothing in those calculi
can observe a unit}. There is no conversion operator, and there cannot be one. A
term that converts meters to feet multiplies by a factor determined by the
magnitudes of the meter and the foot; rescale the meter and the correct
factor changes, so the term would no longer commute with rescaling, and the
unconditional abstraction theorems would be false.

The practical literature has the opposite problem. Practical language designs
and libraries make conversion available: the
object-oriented formulation of
\citet{allen2004}, with its dynamic conversion factors and scales;
F\#'s units~\citep{kennedy2009}, where the programmer writes each
factor as a typed literal; and the metrology library of
\citet{fosterwolff2023}. The gap has
persisted for more than two decades: calculi with theorems have no
conversion, and systems with conversion have no invariance theorems. The
gap is in the theory. In the deployed designs a conversion factor is a
number, whether derived from declared unit magnitudes or written by
hand. What none of them has is a theorem saying what the factor costs
invariance.

In this paper, we close that gap. We present \calc, the simply typed lambda
calculus extended with quantity types indexed by units
($\Q{\mathsf{meter}}$ is the type of lengths measured in meters),
quantification over units and over dimensions, dimensioned vectors and
linear maps, and a typed conversion primitive. The primitive
$\cvt{e}{u}{v}$ takes a
quantity at unit $u$ to the same quantity at unit $v$, multiplying its
magnitude by the appropriate factor; it is well-typed only when $u$ and $v$
share a dimension. The question our paper answers is what survives of the
parametric theory once this term is in the language, and the answer has
three parts.

First, invariance survives, and conversion tells us exactly by how much. We call a
term \emph{parametric} when it names no unit constant $\mathbf{1}_u$
(the constant magnitude $1$ at unit $u$), and
a rescaling \emph{coherent} when it factors through dimension, so that it
cannot separate units of one dimension: if it doubles the meter, it must
double the foot and the yard, all being lengths, which is to say it
leaves the declared factors where they are. A kernel that rescales
one length unit while fixing others makes an incoherent change, outside
this guarantee. Section~\ref{sec:twist} gives a separate certificate for invariance
under independent unit rescalings. Every parametric term,
conversions included, is invariant under every coherent rescaling; every
parametric \emph{convert-free} term is invariant under every rescaling
whatsoever, Kennedy's theorem in our setting. The pair brackets conversion
exactly: a single conversion of a nonzero argument is invariant under a
rescaling if and only if that rescaling assigns its source and target the
same factor, which is what coherence demands of a same-dimension pair.

Second, for first-order programs the theory becomes an algorithm. A
program is \emph{first-order} when its inputs and result are scalars,
vectors, or matrices rather than functions; internal abstraction and
application remain unrestricted. We begin with the scalar case.
We abbreviate meter, foot, and second as
$\mathsf{m}$, $\mathsf{ft}$, and $\mathsf{s}$.
Writing $e\;\mathsf{in}\;v$ for the conversion of $e$ to $v$, and taking
$x$ of type $\Q{\mathsf{m}}$, the round trip
$(x\;\mathsf{in}\;\mathsf{ft})\;\mathsf{in}\;\mathsf{m}$ multiplies
by a factor and its reciprocal and so cannot depend on how the units are
declared, while the one-way $x\;\mathsf{in}\;\mathsf{ft}$ can. We analyze a program's \emph{accumulated conversion ratio} and prove
that, when the analysis assigns a ratio, a first-order program with
nonzero output is invariant under all rescalings exactly when its ratio
is trivial. Triviality is decidable because ratios, like units, are
exponent vectors over $\mathbb{Q}$. The procedure returns one of three
verdicts: it certifies invariance, names the ratio responsible for a
departure, or declines. A decline supplies no invariance verdict; we call
an assigned ratio the program's \emph{drift}. Drift-free does not require conversion freedom, as the round trip shows: the certificate says the
declarations cannot reach the answer, not that the program avoided
conversion.

Third, conversion forces an account of where its factors come from.
Declarations such as $\mathsf{yard} = 3\,\mathsf{foot}$ and
$\mathsf{foot} = 0.3048\,\mathsf{meter}$ can conflict: a mistyped
$\mathsf{yard} = 0.9\,\mathsf{meter}$ makes a direct conversion and a conversion
through feet disagree, a conversion-library bug. Following
\citet{karrloveman1978}, we give declaration sets a consistency
criterion in exact rational arithmetic, necessary and sufficient, and carry the surviving factor by a chain of theorems from
the declaration through the semantics to the multiplication the
evaluator performs with real arithmetic: a yard evaluates to $0.9144$ meters by either
route, as a theorem. An executable, verified declaration checker also
requires the factors to be determined for every same-dimension pair;
consistency alone does not ensure this. It extracts exact factors,
including radicals,\footnote{Declared factors are rational, but because
exponents are rational, a derived factor can be a radical. A
one-dimensional wavefunction amplitude carries $\mathsf{m}^{-1/2}$; with
$\mathsf{nm} = 10^{-9}\,\mathsf{m}$ declared, converting an amplitude
from $\mathsf{nm}^{-1/2}$ to $\mathsf{m}^{-1/2}$ multiplies by
$10^{9/2}$. The checker extracts every factor exactly, as a rational
radicand and an integer root degree.} from the accepted declarations
(Section~\ref{sec:declarations}).

Every theorem in our paper is a theorem of a Lean~4 mechanization we
call \emph{the
artifact},\footnote{\url{https://github.com/ericeallen/lambda-s}} and
the artifact, not this paper, is the source of truth
(Section~\ref{sec:mechanization}); our paper guides the reader to the
ideas. The artifact also runs: the evaluator the theorems are about,
instantiated at floating point, compiles through Lean's C backend to a
native binary whose vector and matrix operations call BLAS
(Accelerate's \textsf{cblas}) on macOS and portable C loops elsewhere. That is
possible because a matrix's units live in its type
(Section~\ref{sec:hart}), so its payload is an array of doubles with
no per-entry tags, which is exactly what BLAS expects. The binary runs
the instrumented evaluator of Section~\ref{sec:dynamics}, so a scalar
still carries its unit tag; the erasure theorem
(Section~\ref{sec:erasure}) says it need not.

\calc{} has no recursion and no boolean type beyond a
compare-and-branch on quantities, but does have vectors
and linear maps with per-component units, following \citet{hart1995};
we sought the smallest calculus that captures the phenomena we study.
Every term works at one fixed size, since \calc{} has neither recursion
nor quantification over size, but at a fixed size it writes numerical
kernels directly: the artifact reads matrix entries and rows, builds
transposes, computes the trace, determinant, and inverse of a dimensioned
endomorphism, and performs a $2\times 2$ Jacobi rotation that
diagonalizes a symmetric matrix exactly (Section~\ref{sec:hart}).
Algorithms that iterate to convergence, such as a general eigensolver or
a singular value decomposition, are not expressible.

Our paper follows the development in order: the calculus and its checker
(Section~\ref{sec:calculus}), declarations (Section~\ref{sec:declarations}),
the evaluator (Section~\ref{sec:dynamics}), the semantics and the price
of conversion (Sections~\ref{sec:semantics}--\ref{sec:twist}), adequacy
and erasure (Section~\ref{sec:erasure}), dimensional analysis and
dimensioned linear algebra (Sections~\ref{sec:pi}--\ref{sec:hart}), the
mechanization (Section~\ref{sec:mechanization}), related work, and the
conclusion.

\section{The Calculus}
\label{sec:calculus}

We present the syntax and statics of \calc. We represent units and
dimensions as \emph{exponent vectors} over $\mathbb{Q}$, and every operation
on them is linear algebra. The unit $\mathsf{kg}\cdot\mathsf{m}/\mathsf{s}^2$
is the vector $(\mathsf{kg}\mapsto 1,\ \mathsf{m}\mapsto 1,\ \mathsf{s}\mapsto -2)$;
multiplication is pointwise addition, the $q$-th power is scalar
multiplication by $q$, and equality is a comparison of finitely many
rationals. There is nothing to normalize: $\mathrm{u}\cdot\mathrm{v}$ and
$\mathrm{v}\cdot\mathrm{u}$ are one vector, exactly as $2+2$ and $4$ are
one number.

Two aspects of this representation carry our paper. First,
\emph{exponents are rational}, not integral as in Kennedy's free abelian
group~\citep{kennedy1997}. Rational exponents appear in
\citet{karrloveman1978} and in the dimension-inference systems of
\citet{wandokeefe1991} and \citet{goubault1994}, where dimensions already
form a vector space over $\mathbb{Q}$. Kennedy's argument for $\mathbb{Z}$ was that
$\mathsf{M}^{1/2}$ makes no physical sense and would suggest revising
the set of base dimensions~\citep[\S1.3]{kennedy1996}. Any finite set of rational
exponents can be made integral by rebasing at a common denominator.
But a library fixes its basis before a client chooses the powers it
needs; rational exponents admit those powers without changing the
library. The group is then \emph{divisible}, so
$\sqrt{e} : \Q{\mathsf{m}^{1/2}}$ for $e : \Q{\mathsf m}$ needs no change of basis, and in the proofs
rank-nullity replaces Smith normal form (Section~\ref{sec:pi}). Quantities of genuinely fractional
dimension, such as the $\mathsf{m}^{-1/2}$ of a one-dimensional
wavefunction amplitude in quantum mechanics,
come for free.\footnote{A probability density on
a line carries $\mathsf{Length}^{-1}$, so that its integral is
dimensionless. The density is the squared modulus of the amplitude,
so the amplitude carries $\mathsf{Length}^{-1/2}$.} Second, \emph{units and dimensions are related
by a homomorphism}. Let $B$ and $D$ be the sets of base units and base
dimensions. A unit system assigns each base unit a dimension,
$\dimOf : B \to \mathbb{Q}^D$, total and fixed, extended linearly.
Nothing restricts a dimension to one unit, which is precisely the
situation conversion exists to serve and that systems tracking only
dimensions cannot express. The two-level design follows the
object-oriented lineage~\citep{allen2004}.

\begin{figure}
\[
\begin{array}{@{}l@{\quad}r@{\;}c@{\;}l@{\qquad}l@{}}
\text{units} & u, v & ::= & 1 \mid b \mid \mathrm{u} \mid u \cdot v \mid
  u / v \mid u^{q} & b \in B;\ \mathrm{u} \text{ a unit variable};\ q \in \mathbb{Q}\\
\text{dimensions} & d & ::= & 1 \mid d_0 \mid \delta \mid d \cdot d \mid
  d / d \mid d^{q} & d_0 \in D;\ \delta \text{ a dimension variable}\\
\text{spaces} & \vec{u}, \vec{w} & ::= & [\,] \mid u, \vec{u} &
  \text{per-component unit lists}\\[4pt]
\text{types} & \tau & ::= & \Q{u} & \text{quantity at unit } u\\
& & \mid & \tau \to \tau & \text{function}\\
& & \mid & \mathsf{Vec}\,\vec{u} & \text{vector with per-component units}\\
& & \mid & \mathsf{Lin}\,\vec{u}\,\vec{w} & \text{linear map}\\
& & \mid & \uall{\mathrm{u}}{d}\tau & \text{unit abstraction, bounded by a dimension}\\
& & \mid & \dall{\delta}\tau & \text{dimension abstraction}\\[2pt]
\text{terms} & e & ::= & x \mid \lambda x{:}\tau.\,e \mid e\;e & \\
& & \mid & q \mid \mathbf{1}_u & \text{rational literal (at unit } 1\text{); unit constant}\\
& & \mid & e \cdot e \mid e / e \mid e + e \mid e^{q} & \text{arithmetic; constant rational power}\\
& & \mid & \log e \mid \exp e & \text{at unit } 1 \text{ only}\\
& & \mid & \ifle{e}{e}{e}{e} & \text{compare and branch}\\
& & \mid & e.i \mid \mathsf{row}_i\,e & \text{component; row}\\
& & \mid & e \odot e \mid e \circ e & \text{map application; composition}\\
& & \mid & \langle\rangle \mid e \mathbin{::} e & \text{vector literals: empty; cons}\\
& & \mid & \langle\rangle_{\vec{u}} \mid e \mathbin{::}_{w} e & \text{matrix literals: rowless at } \vec{u}\text{; row cons at } w\\
& & \mid & \Lambda \mathrm{u}{:}d.\,e \mid e\,[u] & \text{unit abstraction and application}\\
& & \mid & \Lambda\delta.\,e \mid e\,\{d\} & \text{dimension abstraction and application}\\
& & \mid & \cvt{e}{u}{v} & \text{conversion}\\
\end{array}
\]
\caption{Syntax of \calc. Function application
$e\;e$ and linear-map application $e \odot e$ are distinct term forms, as in
the artifact.}
\Description{Grammar of the calculus: the productions for units,
dimensions, spaces, types, and terms, each with a gloss in the right margin.}
\label{fig:syntax}
\end{figure}

The types and terms of \calc{} appear in Figure~\ref{fig:syntax}. A
quantity type $\Q{u}$ classifies scalars at unit $u$;
$\mathsf{Vec}\,\vec{u}$ and $\mathsf{Lin}\,\vec{u}\,\vec{w}$ classify
vectors and linear maps with per-component units, following
\citet{hart1995}. Both carry introduction forms as well as eliminations:
a vector is built by consing scalars onto $\langle\rangle$, each cons
extending the space, and a matrix by rows, where $\langle\rangle_{\vec{u}}$
carries its domain space (so a rowless matrix has a well-defined width)
and a row cons at output unit $w$ requires component $i$ at unit
$w/u_i$. The two quantifiers are the calculus's
answer to what an \emph{unbounded} unit variable means: there is no such
thing. Unit abstraction is always bounded by a dimension, and unbounded
quantification is recovered as $\dall{\delta}\uall{\mathrm{u}}{\delta}\tau$.
The bound is what makes conversion usable in polymorphic code: under
$\Lambda\mathrm{u}{:}\mathsf{Length}$ the body may convert between
$\mathrm{u}$ and $\mathsf{meter}$, because the checker sees they share a
dimension. In particular the generic caster
$\Lambda\delta.\,\Lambda\mathrm{u}{:}\delta.\,\Lambda\mathrm{v}{:}\delta.\,
\lambda x{:}\Q{\mathrm{u}}.\,\cvt{x}{\mathrm{u}}{\mathrm{v}}$ is
well-typed (\artifact{caster}), and the abstraction theorems of
Section~\ref{sec:semantics} explain what it costs.

The arithmetic rules carry the unit discipline (Figure~\ref{fig:typing}).
Multiplication and division are total on units. Addition demands its
operands at \emph{equal} units, not merely interchangeable ones: adding
meters to feet is a type error, and $\mathsf{convert}$ is how the
programmer says it was intended.\footnote{The value of requiring exact
unit matching was first made clear to the author by Guy L. Steele Jr.,
who argued that it helps a programmer manage the imprecisions and
finite bounds of floating-point computation.} Kennedy's thesis proposes
the opposite design as further work for a units-primary language:
typing rules extended so that inference inserts the conversions
automatically, after Thatte's coercive
isomorphisms~\citep[\S8.1]{kennedy1996}. Powers at constant
rational exponents are primitive, $(\cdot)^q : \Q{u} \to \Q{u^q}$ with no
side condition, and $\sqrt[n]{e}$ abbreviates $e^{1/n}$; not every power
is definable from the field operations, which is the non-definability
theorem of Section~\ref{sec:pi}. Finally, $\log$ and $\exp$ require
arguments at unit $1$. A dimensionless ratio such as $\mathsf m/\mathsf{ft}$
must first be converted to that unit.
Conversion carries its source unit and its target, because the
evaluator uses those units to look up the factor in its supplied conversion
function, without consulting a typing derivation;
we support a surface syntax such as 
$e\;\mathsf{in}\;v$ that leaves the source to the verified checker
(\artifact{elabConvert}). Elaboration succeeds exactly when the
argument has some unit of $v$'s dimension
(\artifact{elabConvert_isSome}).

\begin{figure}
\begin{mathpar}
\inferrule[T-Var]{\Gamma(x) = \tau}{\Delta;\Gamma \vdash x : \tau}
\and
\inferrule[T-Lam]{\Delta;\Gamma, x{:}\tau \vdash e : \sigma}
  {\Delta;\Gamma \vdash \lambda x{:}\tau.\,e : \tau \to \sigma}
\and
\inferrule[T-App]{\Delta;\Gamma \vdash e_1 : \tau \to \sigma \\
  \Delta;\Gamma \vdash e_2 : \tau}{\Delta;\Gamma \vdash e_1\;e_2 : \sigma}
\and
\inferrule[T-Lit]{ }{\Delta;\Gamma \vdash q : \Q{1}}
\and
\inferrule[T-Con]{ }{\Delta;\Gamma \vdash \mathbf{1}_u : \Q{u}}
\and
\inferrule[T-Mul]{\Delta;\Gamma \vdash e_1 : \Q{u} \\
  \Delta;\Gamma \vdash e_2 : \Q{v}}
  {\Delta;\Gamma \vdash e_1 \cdot e_2 : \Q{u\,v}}
\and
\inferrule[T-Div]{\Delta;\Gamma \vdash e_1 : \Q{u} \\
  \Delta;\Gamma \vdash e_2 : \Q{v}}
  {\Delta;\Gamma \vdash e_1 / e_2 : \Q{u/v}}
\and
\inferrule[T-Add]{\Delta;\Gamma \vdash e_1 : \Q{u} \\
  \Delta;\Gamma \vdash e_2 : \Q{u}}
  {\Delta;\Gamma \vdash e_1 + e_2 : \Q{u}}
\and
\inferrule[T-IfLe]{\Delta;\Gamma \vdash e_1 : \Q{u} \\
  \Delta;\Gamma \vdash e_2 : \Q{u} \\
  \Delta;\Gamma \vdash e_3 : \tau \\ \Delta;\Gamma \vdash e_4 : \tau}
  {\Delta;\Gamma \vdash \ifle{e_1}{e_2}{e_3}{e_4} : \tau}
\and
\inferrule[T-Pow]{\Delta;\Gamma \vdash e : \Q{u}}
  {\Delta;\Gamma \vdash e^{q} : \Q{u^{q}}}
\and
\inferrule[T-VNil]{ }{\Delta;\Gamma \vdash \langle\rangle : \mathsf{Vec}\,[\,]}
\and
\inferrule[T-VCons]{\Delta;\Gamma \vdash e_1 : \Q{u} \\
  \Delta;\Gamma \vdash e_2 : \mathsf{Vec}\,\vec{u}}
  {\Delta;\Gamma \vdash e_1 \mathbin{::} e_2 : \mathsf{Vec}\,(u, \vec{u})}
\and
\inferrule[T-MNil]{ }{\Delta;\Gamma \vdash \langle\rangle_{\vec{u}} :
  \mathsf{Lin}\,\vec{u}\,[\,]}
\and
\inferrule[T-MCons]{\Delta;\Gamma \vdash e_1 : \mathsf{Vec}\,(w/\vec{u}) \\
  \Delta;\Gamma \vdash e_2 : \mathsf{Lin}\,\vec{u}\,\vec{w}}
  {\Delta;\Gamma \vdash e_1 \mathbin{::}_{w} e_2 :
  \mathsf{Lin}\,\vec{u}\,(w, \vec{w})}
\and
\inferrule[T-Idx]{\Delta;\Gamma \vdash e : \mathsf{Vec}\,\vec{u} \\
  \vec{u}_i = u}{\Delta;\Gamma \vdash e.i : \Q{u}}
\and
\inferrule[T-MRow]{\Delta;\Gamma \vdash e : \mathsf{Lin}\,\vec{u}\,\vec{w} \\
  \vec{w}_j = w}{\Delta;\Gamma \vdash \mathsf{row}_j\,e : \mathsf{Vec}\,(w/\vec{u})}
\and
\inferrule[T-Mapp]{\Delta;\Gamma \vdash e_1 : \mathsf{Lin}\,\vec{u}\,\vec{w} \\
  \Delta;\Gamma \vdash e_2 : \mathsf{Vec}\,\vec{u}}
  {\Delta;\Gamma \vdash e_1 \odot e_2 : \mathsf{Vec}\,\vec{w}}
\and
\inferrule[T-Comp]{\Delta;\Gamma \vdash e_1 : \mathsf{Lin}\,\vec{v}\,\vec{w} \\
  \Delta;\Gamma \vdash e_2 : \mathsf{Lin}\,\vec{u}\,\vec{v}}
  {\Delta;\Gamma \vdash e_1 \circ e_2 : \mathsf{Lin}\,\vec{u}\,\vec{w}}
\and
\inferrule[T-Log]{\Delta;\Gamma \vdash e : \Q{1}}
  {\Delta;\Gamma \vdash \log e : \Q{1}}
\and
\inferrule[T-Exp]{\Delta;\Gamma \vdash e : \Q{1}}
  {\Delta;\Gamma \vdash \exp e : \Q{1}}
\and
\inferrule[T-ULam]{\Delta, \mathrm{u}{:}d;\Gamma \vdash e : \tau}
  {\Delta;\Gamma \vdash \Lambda\mathrm{u}{:}d.\,e : \uall{\mathrm{u}}{d}\tau}
\and
\inferrule[T-UApp]{\Delta;\Gamma \vdash e : \uall{\mathrm{u}}{d}\tau \\
  \dimOf_\Delta(w) = d}
  {\Delta;\Gamma \vdash e\,[w] : \tau[\mathrm{u} \mapsto w]}
\and
\inferrule[T-DLam]{\Delta, \delta;\Gamma \vdash e : \tau}
  {\Delta;\Gamma \vdash \Lambda\delta.\,e : \dall{\delta}\tau}
\and
\inferrule[T-DApp]{\Delta;\Gamma \vdash e : \dall{\delta}\tau}
  {\Delta;\Gamma \vdash e\,\{d\} : \tau[\delta \mapsto d]}
\and
\inferrule[T-Cvt]{\Delta;\Gamma \vdash e : \Q{u} \\ u \sim_\Delta v}
  {\Delta;\Gamma \vdash \cvt{e}{u}{v} : \Q{v}}
\end{mathpar}
\caption{Typing: the full rule set, transcribed from the artifact's
\protect\artifact{HasTy} with named variables in place of its de Bruijn
indices. $\Delta;\Gamma \vdash e : \tau$ types $e$ under
a context $\Delta$ of dimension variables and unit variables (each with its
dimension bound) and a term context $\Gamma$; $\dimOf_\Delta$ extends $\dimOf$ to unit
variables by their bounds, and $u \sim_\Delta v$ means
$\dimOf_\Delta(u) = \dimOf_\Delta(v)$.
In \ftrule{T-ULam} and \ftrule{T-DLam} the bound variable is fresh for
$\Delta$, so it occurs nowhere in $\Gamma$. In \ftrule{T-Idx} and \ftrule{T-MRow}, the premises
$\vec{u}_i = u$ and $\vec{w}_j = w$ abbreviate successful bounds-checked
lookups, as in the artifact. In \ftrule{T-Con}, $u$ ranges over all unit expressions,
variables included. $(u, \vec{u})$ prepends a component to a space, and in
\ftrule{T-MCons}, $w/\vec{u}$ is the pointwise quotient, entry $j$ at
$w/u_j$.}
\Description{The typing rules of the calculus as inference rules, one
per term form, with bracketed small-capital names.}
\label{fig:typing}
\end{figure}

The statics borrow a practice from dependently typed
programming~\citep{cpdt}: the checker returns not a type but a
\emph{derivation}, an inhabitant of
$\Sigma\tau.\,\mathrm{HasTy}\,\Delta\,\Gamma\,e\,\tau$. Checker
soundness (anything it accepts is well-typed) is then a consequence of
the type rather than a theorem: a checker that cannot produce a type
without the derivation justifying it cannot accept an ill-typed term.
(Type soundness, about the dynamics, is proved in
Section~\ref{sec:dynamics}.) What
remains with content is completeness, which the artifact proves in a
strengthened form: the checker returns a derivation whenever one exists,
and derivations are unique, so we can define the semantics by
recursion on them (\artifact{check_eq}, which gives uniqueness as a corollary).

The checker consumes fully annotated terms; it is not a principal-type
inference algorithm. Kennedy's inference solves unit equations by
abelian-group unification~\citep{kennedy1996}. Our module
\artifact{LambdaS.Unify} proves single-equation elimination, preservation
of ground solution sets by terminating triangularization
(\artifact{System.solves_triangulate_iff}), and the terminal check that
the variable-free equations left after triangularization already hold.
It does not construct a most general symbolic substitution or generate
constraints from unannotated terms. Conversion also contributes
$\dimOf\ u=\dimOf\ v$, which does not imply $u=v$. Whether decidable
principal-type inference extends to this two-level constraint system
remains open in our development.

\section{Unit Declarations}
\label{sec:declarations}

Conversion needs factors, and factors need a source and a soundness
criterion. In \calc{} they come from declarations:
\[
\mathsf{unit}\;\mathsf{yard} = 3\,\mathsf{foot}; \qquad
\mathsf{unit}\;\mathsf{foot} = 0.3048\,\mathsf{meter}.
\]
Declarations can conflict: add a mistyped
$\mathsf{yard} = 0.9\,\mathsf{meter}$ and the direct route disagrees
with the route through feet, since $3 \times 0.3048 = 0.9144 \neq
0.9$.\footnote{The yard has been exactly $0.9144$ meters only since
1959, when the United States and five other English-speaking countries
adopted a common yard, the American one moving by two parts per
million~\citep{astinkaro1959}. The United States
kept its earlier foot for surveying; that redundant declaration,
inconsistent with the new one in the seventh decimal place, survived
until its retirement, effective December 31, 2022, announced by a 2020
Federal Register notice~\longcitep{National Institute of Standards and Technology and
National Oceanic and Atmospheric Administration}{surveyfoot2020}.}
An implementation that chooses a conversion route without first
checking consistency can admit this conversion-library bug. \calc{} chooses
no route. A \emph{valuation} $V$ assigns each base unit a positive real
magnitude (an exchange-rate table, not a measurement), extended
homomorphically to all algebraically expressible units.
Conversion factors are ratios of magnitudes, so path independence is a theorem,
and a declaration set is \emph{consistent} when some valuation satisfies
every declared equation.

The other declaration forms need only scoping:
a bare $\mathsf{dimension}\;\mathsf{Length}$ supplies the parameter $D$;
an abbreviation $\mathsf{dimension}\;\mathsf{Velocity} =
\mathsf{Length}/\mathsf{Time}$ expands at once, so cycles are
unrepresentable rather than detected (a cycle would need a second
binding of $\mathsf{Length}$, rejected for rebinding a generator,
\artifact{dimCycle}; elaboration accepts exactly the sequences whose
references point backwards, \artifact{DimAbbrev.elabDimDefs_isSome_iff}); and a \emph{primary unit}
declaration (the term is \citeauthor{allen2004}'s
\citeyearpar{allen2004}, the syntax Fortress's~\citep{fortress2006}),
$\mathsf{unit}\;\mathsf{meter} : \mathsf{Length}$, supplies $B$ and
$\dimOf$ (\artifact{elabPrimary}), needing only distinct names and
dimensions that denote (\artifact{DimAbbrev.elabPrimary_isSome_iff}). Factor declarations, by contrast,
carry semantic content, and cycles among them are just dependencies: the
benign $\mathsf{foot} = \tfrac{1}{3}\,\mathsf{yard}$ is satisfiable
(\artifact{cycle_satisfiable}); $\mathsf{foot} = \mathsf{yard}$ forces
$3 = 1$ and is rejected (\artifact{cycle_conflict}).

Each factor declaration $u = q\,v$ is a ratio $r = u/v$ in the unit
group and a positive rational factor $q$, and a valuation satisfies it exactly when
$V(r) = q$. The mistyped set has $r_1 = \mathsf{yard}/\mathsf{foot}$,
$r_2 = \mathsf{foot}/\mathsf{meter}$, $r_3 = \mathsf{yard}/\mathsf{meter}$
with $q_1 = 3$, $q_2 = 0.3048$, $q_3 = 0.9$. Its ratios are dependent,
$r_1 \cdot r_2 = r_3$, that is, $r_1 + r_2 - r_3 = 0$ as exponent vectors,
so any valuation must have $q_1 q_2/q_3 = V(r_1)V(r_2)/V(r_3) = V(1) = 1$;
but $3 \times 0.3048/0.9 = 1.016$, so none exists. With $q_3 = 0.9144$
the product is exactly $1$. The three ratios have rank $2$, so this is
their only dependency, and $V(\mathsf{meter}) = 1$, $V(\mathsf{foot}) = 0.3048$,
$V(\mathsf{yard}) = 0.9144$ satisfies all three declarations. The general statement
is this computation over every dependency.

\begin{theorem}[Consistency; \artifact{consistent_iff_dependencies_mul}]
\label{thm:consistency}
A finite set of declarations $u_i = q_i\,v_i$, each of a base unit $u_i$
against an arbitrary unit expression $v_i$, with ratios $r_i = u_i/v_i$
and positive rational factors $q_i$, admits a
satisfying valuation if and only if every $\mathbb{Q}$-linear
combination with $\sum_i c_i\,r_i = 0$ in the unit group has
$\prod_i q_i^{\,c_i} = 1$.
\end{theorem}

In logarithmic coordinates each declaration is one linear equation,
as \citet{karrloveman1978} observed. The artifact solves this system
by executable Gaussian elimination and back-substitution
(\artifact{DeclSolver.solve_isSome_iff}). Coefficients are rational;
the right-hand sides are symbolic rational combinations of logarithms of
positive rationals. For example, $(\log 2)/2$ is stored as the logarithm
of the rational $2$ divided by the integer $2$, without evaluating it.
Clearing denominators expresses such a combination
as $(\log q)/N$, for positive rational $q$ and positive integer $N$.
It is zero exactly when $q=1$, an exact rational test. Thus elimination
never evaluates a real logarithm or root, and a nonzero right-hand side
on a zero row rejects an inconsistent set. Soundness and completeness
connect success to existence of a real satisfying valuation.

Rational powers can require irrational magnitudes. A valuation
satisfying $\mathsf{nm}=10^{-9}\,\mathsf{m}$ with $V(\mathsf{m})=1$ has
$V(\mathsf{nm}^{-1/2})=10^{9/2}$. We therefore take valuations over the
reals. For sufficiency in Theorem~\ref{thm:consistency}, the unknowns
$\log V(b)$ are real, yet checking
only rational dependencies suffices because $\mathbb{R}$ is a vector
space over $\mathbb{Q}$.
The compatible assignment of log-factors defines a $\mathbb{Q}$-linear
map on the span of the ratio vectors; extending it to the whole unit
space gives a satisfying valuation (\artifact{dependency_sufficient}). The mistyped set
is rejected outright: no valuation exists, so there is no route to
choose.

Before execution, every conversion the program can perform must have a
supplied numerical factor. Consistency alone does not ensure this.
Declaring $\mathsf{meter}$ and $\mathsf{foot}$ as primary units of length
makes $x\;\mathsf{in}\;\mathsf{meter}$ well-typed for
$x:\Q{\mathsf{foot}}$, but supplies no numerical answer. The artifact's
declaration checker rejects this incomplete system; adding
$\mathsf{foot}=0.3048\,\mathsf{meter}$ makes it pass.

For a consistent set, $V(u)/V(v)$ is fixed across its satisfying valuations
exactly when the exponent vector $u/v$ lies in the $\mathbb{Q}$-span of the
declared ratios (\artifact{Decl.determined_iff_coefficients}). On that
span linearity fixes the log-factor; outside it a separating linear
functional changes the factor while preserving every declaration.
The executable factor lookup solves for the span coefficients and
succeeds exactly when the factor is determined
(\artifact{DeclSolver.conversion_isSome_iff}). To cover every pair of
same-dimension units, the declared ratios must span the kernel of
$\dimOf$. The global check decides this finite linear-algebra condition,
with both directions proved
(\artifact{DeclarationComplete.check_iff_all_conversions_determined}).

\begin{theorem}[Executable declaration checking; \artifact{DeclSolver.check_isSome_iff}]
For a finite unit system, the declaration checker accepts exactly when
every declaration relates units of the same dimension, a satisfying
valuation exists, and every same-dimension conversion has the same factor
in every satisfying valuation.
\end{theorem}

Acceptance returns a computed solution with proofs of these properties.
Factor extraction then succeeds for every same-dimension pair of concrete
units (\artifact{DeclSolver.Checked.conversionExact_isSome}). It returns
a positive rational radicand and a positive integer root degree, and the
factor they represent is $V(u)/V(v)$ in every satisfying valuation
(\artifact{DeclSolver.conversionExact_correct}).
For example, with $\mathsf{nm}=10^{-9}\,\mathsf{m}$ declared, converting a
wavefunction amplitude from $\mathsf{nm}^{-1/2}$ to $\mathsf{m}^{-1/2}$
multiplies by $10^{9/2}$, returned as radicand $10^{9}$ and degree $2$
(\artifact{Nanometer.exactAmplitude}).
The solver may choose reference magnitudes for unconstrained directions,
but it never uses that choice to claim a missing factor is determined.
Path independence holds within a valuation; the determinacy criterion
above gives agreement across satisfying valuations. Section~\ref{sec:twist}'s
certificate of independence from every supplied factor is stronger
(\artifact{den_indep_of_driftFree}), while the declaration checker
ensures that the conversions needed to execute the program have factors.

\section{Dynamics}
\label{sec:dynamics}

We now run the calculus, with a total conversion function supplied as an
input. Section~\ref{sec:declarations}'s checker and factor extractor
establish that accepted declarations supply every same-dimension factor.
We write $E=(\rho,\eta,\theta)$ for the term, unit, and dimension
environments. The latter two assign a ground unit or dimension to every
variable in scope, because the only rule that brings one into scope,
\ftrule{E-UApp} or \ftrule{E-DApp}, grounds its argument in the caller's
environment before binding it, and evaluation starts on a closed term
with empty environments. Write $u^E$ and $d^E$ for their linear
extensions to unit and dimension expressions, and $\vec u^E$ for
componentwise grounding; these are always closed, and no run-time value
carries a unit variable. Magnitudes range over a carrier $R$ with the artifact's
\artifact{Num} operations. Values are scalars, vectors, matrices, or
closures of three kinds, each capturing all three environments
(Figure~\ref{fig:dyn-closures}).

The evaluator is a definitional interpreter~\citep{reynolds1972}
written as a total function. Lean requires every function to terminate,
and termination of evaluation is a theorem proved below rather than a
fact available at the definition, so the evaluator takes a natural
number, its \emph{fuel}, and returns failure when the fuel runs
out~\citep{aminrompf2017,owens2016}. Intuitively, fuel is a finite
resource like stack space, supplied in whatever quantity a run needs: a
run that completes within the bound returns the same value under every
larger bound (\artifact{eval_mono}), and a run that exhausts it fails
rather than diverging. The judgment
$E;n\vdash e\Downarrow v$ means that \artifact{eval}, with conversion
function $c$ and fuel $n$, returns $v$. We hold $c$ fixed in the rules. The environment, fuel, and term are
inputs; the value is the output. Figures~\ref{fig:dyn-closures}--\ref{fig:dyn-arrays}
give every successful case of the evaluator. A term with no applicable
rule returns failure, including every closure application at fuel zero.
Failure thus represents either exhausted fuel or a failed dynamic check;
the evaluator does not distinguish them. Fuel bounds closure-entry depth,
not a global count of steps: subterms receive the same fuel as their
parent, and only entering a closure body decreases it.

\begin{figure}[htbp]
\small
\[
\begin{array}{rcl}
v & ::= & \langle a,u\rangle
 \mid \mathsf{vec}(\vec a,U)
 \mid \mathsf{mat}(M,U,W)\\
&&{}\mid \mathsf{clos}(x{:}\tau,e,E)
 \mid \mathsf{uclos}(\mathrm{u}{:}d,e,E)
 \mid \mathsf{dclos}(\delta,e,E).
\end{array}
\]
\begin{mathpar}
\inferrule[E-Var]{\rho(x)=v}
 {E;n\vdash x\Downarrow v}
\and
\inferrule[E-Lam]{ }
 {E;n\vdash \lambda x{:}\tau.\,e\Downarrow\mathsf{clos}(x{:}\tau,e,E)}
\and
\inferrule[E-ULam]{ }
 {E;n\vdash \Lambda\mathrm u{:}d.\,e\Downarrow\mathsf{uclos}(\mathrm u{:}d,e,E)}
\and
\inferrule[E-DLam]{ }
 {E;n\vdash\Lambda\delta.\,e\Downarrow\mathsf{dclos}(\delta,e,E)}
\and
\inferrule[E-App]{
 E;n+1\vdash f\Downarrow\mathsf{clos}(x{:}\tau,e,(\rho',\eta',\theta'))\\
 E;n+1\vdash a\Downarrow v\\
 (\rho'[x\mapsto v],\eta',\theta');n\vdash e\Downarrow w}
 {E;n+1\vdash f\;a\Downarrow w}
\and
\inferrule[E-UApp]{
 E;n+1\vdash f\Downarrow\mathsf{uclos}(\mathrm u{:}d,e,(\rho',\eta',\theta'))\\
 (\rho',\eta'[\mathrm u\mapsto u^E],\theta');n\vdash e\Downarrow v}
 {E;n+1\vdash f\,[u]\Downarrow v}
\and
\inferrule[E-DApp]{
 E;n+1\vdash f\Downarrow\mathsf{dclos}(\delta,e,(\rho',\eta',\theta'))\\
 (\rho',\eta',\theta'[\delta\mapsto d^E]);n\vdash e\Downarrow v}
 {E;n+1\vdash f\,\{d\}\Downarrow v}
\end{mathpar}
\caption{Instrumented dynamics: values, variables, and closures.
A scalar carries its ground unit $u$; a vector carries its space $U$;
a matrix carries its domain $U$, codomain $W$, and magnitude rows $M$.
The notation $\rho(x)=v$ denotes a successful lookup, and
$\rho[x\mapsto v]$ extends an environment. Unit and dimension arguments
are grounded in the caller's environment before extending the captured
one. The stored type and dimension annotations impose no dynamic check.}
\Description{Evaluation rules for variables, three closure constructors,
and their applications, showing the decrease in fuel on closure entry.}
\label{fig:dyn-closures}
\end{figure}

\begin{figure}[htbp]
\small
\begin{mathpar}
\inferrule[E-Lit]{ }
 {E;n\vdash q\Downarrow\langle\operatorname{rat}(q),1\rangle}
\and
\inferrule[E-Con]{ }
 {E;n\vdash\mathbf1_u\Downarrow\langle\operatorname{rat}(1),u^E\rangle}
\and
\inferrule[E-Mul]{
 E;n\vdash e_1\Downarrow\langle a,u\rangle\\
 E;n\vdash e_2\Downarrow\langle b,v\rangle}
 {E;n\vdash e_1\cdot e_2\Downarrow\langle a\times_R b,uv\rangle}
\and
\inferrule[E-Div]{
 E;n\vdash e_1\Downarrow\langle a,u\rangle\\
 E;n\vdash e_2\Downarrow\langle b,v\rangle}
 {E;n\vdash e_1/e_2\Downarrow\langle a\div_R b,u/v\rangle}
\and
\inferrule[E-Add]{
 E;n\vdash e_1\Downarrow\langle a,u\rangle\\
 E;n\vdash e_2\Downarrow\langle b,u\rangle}
 {E;n\vdash e_1+e_2\Downarrow\langle a+_R b,u\rangle}
\and
\inferrule[E-Pow]{E;n\vdash e\Downarrow\langle a,u\rangle}
 {E;n\vdash e^q\Downarrow\langle\operatorname{pow}_R(a,q),u^q\rangle}
\and
\inferrule[E-Trans]{E;n\vdash e\Downarrow\langle a,1\rangle\\
 h\in\{\log,\exp\}}
 {E;n\vdash h\,e\Downarrow\langle h_R(a),1\rangle}
\and
\inferrule[E-If]{
 E;n\vdash e_1\Downarrow\langle a,u\rangle\\
 E;n\vdash e_2\Downarrow\langle b,u\rangle\\
 \beta=\operatorname{le}(a,b)\\
 E;n\vdash e_\beta\Downarrow v}
 {E;n\vdash\ifle{e_1}{e_2}{e_{\mathsf{true}}}{e_{\mathsf{false}}}\Downarrow v}
\and
\inferrule[E-Cvt]{
 E;n\vdash e\Downarrow\langle a,u^E\rangle\\
 \dimOf(u^E)=\dimOf(v^E)}
 {E;n\vdash\cvt{e}{u}{v}\Downarrow
   \langle a\mathbin{\times_R}c(u^E,v^E),v^E\rangle}
\end{mathpar}
\caption{Instrumented dynamics: scalars and branching. The transcendental
schema covers logarithm and exponential. All magnitude operations are
those supplied by \protect\artifact{Num}: $\operatorname{rat}$,
$\times_R$, $\div_R$, $+_R$, $\operatorname{pow}_R$ (base first; the
artifact's \protect\artifact{Num} takes the exponent first), $\log_R$, $\exp_R$,
and the Boolean comparison $\operatorname{le}$. No positivity or
nonzero-divisor check is added. Only the selected branch is evaluated.}
\Description{Rules for scalar constants, arithmetic, powers,
logarithm and exponential, compare-and-branch, and conversion.}
\label{fig:dyn-scalars}
\end{figure}

\begin{figure}[htbp]
\small
\begin{mathpar}
\inferrule[E-VNil]{ }
 {E;n\vdash\langle\rangle\Downarrow\mathsf{vec}([\,],[\,])}
\and
\inferrule[E-VCons]{
 E;n\vdash e\Downarrow\langle a,u\rangle\\
 E;n\vdash v\Downarrow\mathsf{vec}(\vec a,U)}
 {E;n\vdash e::v\Downarrow\mathsf{vec}(a::\vec a,u::U)}
\and
\inferrule[E-MNil]{ }
 {E;n\vdash\langle\rangle_{\vec u}\Downarrow
   \mathsf{mat}([\,],\vec u^E,[\,])}
\and
\inferrule[E-MCons]{
 E;n\vdash r\Downarrow\mathsf{vec}(\vec a,Z)\\
 E;n\vdash m\Downarrow\mathsf{mat}(M,U,W)}
 {E;n\vdash r::_{w}m\Downarrow\mathsf{mat}(\vec a::M,U,w^E::W)}
\and
\inferrule[E-Idx]{
 E;n\vdash e\Downarrow\mathsf{vec}(\vec a,U)\\
 \vec a[i]=a\quad U[i]=u}
 {E;n\vdash e.i\Downarrow\langle a,u\rangle}
\and
\inferrule[E-Row]{
 E;n\vdash e\Downarrow\mathsf{mat}(M,U,W)\\
 M[i]=\vec a\quad W[i]=w}
 {E;n\vdash\mathsf{row}_i\,e\Downarrow\mathsf{vec}(\vec a,w/U)}
\and
\inferrule[E-MApp]{
 E;n\vdash f\Downarrow\mathsf{mat}(M,U,W)\\
 E;n\vdash x\Downarrow\mathsf{vec}(\vec a,U)}
 {E;n\vdash f\odot x\Downarrow\mathsf{vec}(\operatorname{matVec}(M,\vec a),W)}
\and
\inferrule[E-Comp]{
 E;n\vdash f\Downarrow\mathsf{mat}(M,V,W)\\
 E;n\vdash g\Downarrow\mathsf{mat}(N,U,V)}
 {E;n\vdash f\circ g\Downarrow\mathsf{mat}(M\mathbin{\odot_{|U|}}N,U,W)}
\end{mathpar}
\caption{Instrumented dynamics: vectors and matrices. All indexed
lookups must succeed. The space $w/U$ has component $i$ at $w/u_i$.
The operation $\operatorname{matVec}$ is supplied by
\protect\artifact{Num}. For each row $r$ of $M$, the product
$M\mathbin{\odot_p}N$ constructs the row
$[\operatorname{dot}(r,\operatorname{col}_i(N))]_{i=0}^{p-1}$ using the carrier's
$\operatorname{dot}$; $\operatorname{col}_i$ replaces a missing
entry in a row by $\operatorname{rat}(0)$. Matrix row construction ignores
$Z$, the incoming row's unit space. No rule checks payload lengths beyond
the explicit indexed lookups; typing and soundness establish the shapes
for well-typed executions.}
\Description{Evaluation rules for empty and nonempty vectors and
matrices, indexing, row projection, map application, and composition.}
\label{fig:dyn-arrays}
\end{figure}

The evaluator checks units wherever an operation is meaningful only
at particular units, and nowhere else. Addition, comparison,
conversion, map application, and composition each check that their two
operands agree; logarithm and exponential check that their one operand
is dimensionless; multiplication, division, and powers compute their
result unit and check nothing; and the introduction forms record their
annotations without inspecting their arguments. The checks are redundant for well-typed terms, and that
redundancy is Theorem~\ref{thm:unit-soundness} below: the stuck states
are unreachable. The one constructor where this matters is
$\mathsf{mcons}$, which records its annotated output unit and discards
the row's units, since a matrix value stores bare magnitudes. An
ill-typed row would corrupt every later product undetected; the typing
rule alone excludes it, and soundness carries that guarantee to the
eliminators.

The soundness and normalization results are parameterized by the
supplied conversion function. The fuel is not assumed but
produced: every closed well-typed term evaluates at some fuel, by a
reducibility argument in the style of
\citet{tait1967}\footnote{Quantification ranges over units and dimensions,
first-order algebraic data, so there is no impredicativity and no need
for Girard's candidates~\citep{girard1972}.}
(\artifact{eval_terminates}).
The calculus has no reduction relation; what is proved is termination of
the evaluator, which is the normalization the fuel needs. Type
soundness then has its usual two halves: preservation
(\artifact{eval_sound}), and progress in the form the fueled setting
delivers, that enough fuel always exists (\artifact{eval_total}).

\begin{theorem}[Unit soundness; \artifact{unit_soundness_total}]
\label{thm:unit-soundness}
For every closed well-typed term $e : \Q{u}$ there exists a fuel bound
$n$ and a magnitude $m$ such that evaluation of $e$ at fuel $n$ yields
exactly $\langle m, u\rangle$.
\end{theorem}

The unit in the conclusion is $u$ itself: the run-time tag provably
agrees with the static type, which is what makes the tag erasable
(Section~\ref{sec:erasure}). Fuel is observable: in the artifact's
quantum system example (\artifact{LambdaS.QM}), a scalar computation
written as an open term over its inputs evaluates at fuel zero, and,
curried and reapplied, it fails there and needs fuel one
(\artifact{twoStateChecks}).

\section{The Price of Conversion}
\label{sec:semantics}

This section states what conversion costs. The semantics rests on one
decision, and it is where our development parts from prior
formalizations: a unit abstraction denotes a \emph{family} of meanings,
indexed by the magnitude the bound unit receives. A calculus that cannot
observe units may interpret the family as constant, which is why the
mechanized semantics of \citet{kennedy2008} could interpret an
abstraction as just its body (``units ignored''); conversion is exactly
what forces the family to vary. A \emph{rescaling} $\psi$ assigns each base
unit and unit variable a positive factor, extended homomorphically; it is
\emph{coherent} when it factors through dimension, so that
units of one dimension move together. A coherent rescaling therefore
leaves every same-dimension conversion factor fixed, and an incoherent
one changes at least one such factor. We call a term \emph{parametric}
when it contains no unit constant $\mathbf{1}_u$.

There are two independent actions. Rescaling inputs by $\psi$ multiplies
an argument at $\Q{u}$ by $\psi(u)$. With the valuation fixed, doubling
a meter input doubles $x\;\mathsf{in}\;\mathsf{ft}$. Rescaling the
valuation by $\varphi$ instead replaces $V(b)$ by $V(b)\varphi(b)$. The
valuation moves with the inputs, not against them: a rescaling is a
change of representation applied to the stored numerals and to the
declared table alike, not a change in the size of a physical unit,
under which numerals and magnitudes would move inversely and every
conversion would be trivially invariant.
With the input fixed, doubling the meter's valuation and fixing the
foot's doubles the conversion factor $V(\mathsf m)/V(\mathsf{ft})$.
Doing both therefore multiplies the answer by four, although its type
$\Q{\mathsf{ft}}$ predicts no change. The abstraction theorems below
specialize these actions to $\varphi=\psi$; Section~\ref{sec:twist}
states the law with the actions separate.

Unit application also consults the valuation to select a member of a
family; conversion uses it to multiply a magnitude. Convert-free terms
nevertheless have valuation-independent denotations in independent
environments, at every type, and at scalar type this is equality of the
two numbers (\artifact{den_indep}). This is a semantic independence
property, not an absence of valuation lookups.

The proofs use two logical relations, $\mathcal R_\psi$
(\artifact{Rel}) and $\mathcal R^{\mathrm{co}}_{\Delta,\Phi,\psi}$
(\artifact{RelCo}), where $\Delta$ assigns dimensions to unit variables
and $\Phi$ rescales dimensions. They agree at quantifier-free types.
At $\Q{u}$ both relate $x$ to $\psi(u)\cdot x$; at an arrow they
map related arguments to related results. Vectors are related
componentwise; entry $(j,i)$ of a linear map scales by
$\psi(w_j)/\psi(u_i)$, as in Hart's rank-one factoring
(Section~\ref{sec:hart}).

The difference is at a unit quantifier $\uall{\mathrm{u}}{d}\tau$.
Write $F(r)$ and $G(r)$ for the values of two families when the bound
unit has log-magnitude $r$, and $\psi[\mathrm{u}\mapsto e^s]$ for
extending a rescaling.
The convert-free relation requires
\[
  \forall r,s\in\mathbb R.\quad
  \mathcal R_{\psi[\mathrm{u}\mapsto e^s]}^{\tau}
       (F(r),G(r+s)).
\]
The coherent relation instead requires
\[
  \forall r\in\mathbb R.\quad
  \mathcal R^{\mathrm{co},\tau}_{\Delta,\mathrm{u}:d,\Phi,
       \psi[\mathrm{u}\mapsto\Phi(d)]}
       (F(r),G(r+\log\Phi(d))).
\]
The family's index remains quantified, but the bound unit's factor is
\emph{determined} by its dimension. Quantifying that factor independently
would let a bound length rescale differently from the meter, invalidating
conversion to meters under the binder. At a dimension quantifier,
$\mathcal R$ uses the body's relation unchanged, whereas
$\mathcal R^{\mathrm{co}}$ quantifies over every positive factor for the
new dimension, extending $\Phi$ accordingly. Theorem~\ref{thm:abs-free}
below is therefore not the special case of
Theorem~\ref{thm:abs-coherent} at a coherent $\psi$: at a unit binder it
asks for relatedness at every factor for the bound unit, where the
coherent relation fixes one.

Conversion $\cvt{x}{u}{v}$ multiplies by
$\mathrm{conv}_V(u,v)=V(u)/V(v)$, so relatedness at a nonzero argument
requires $\psi(u)=\psi(v)$. The fundamental lemmas prove relatedness of
denotations at $V$ and $V\psi$, in related environments: with
$\mathcal R$ for convert-free terms, and with $\mathcal R^{\mathrm{co}}$
when $\psi$ factors through $\Phi$.

\begin{theorem}[Abstraction, convert-free; \artifact{fundamental_free}]
\label{thm:abs-free}
Every parametric, convert-free term preserves $\mathcal R_\psi$-related
environments under every rescaling $\psi$: its denotations at $V$ and
$V\psi$ are $\mathcal R_\psi$-related at its result type.
\end{theorem}

\begin{theorem}[Abstraction, coherent; \artifact{fundamental}]
\label{thm:abs-coherent}
Every parametric term, conversions included, preserves
$\mathcal R^{\mathrm{co}}_{\Delta,\Phi,\psi}$-related environments whenever
$\psi$ factors through $\Phi$: its denotations at $V$ and $V\psi$ are
related by $\mathcal R^{\mathrm{co}}_{\Delta,\Phi,\psi}$ at its result type.
\end{theorem}

Theorem~\ref{thm:abs-free} is Kennedy's
theorem~\citeyearpar{kennedy1997}, here with vectors, linear maps, both
quantifiers, and rational powers; we call its equation a term's
\emph{scaling law}. His scalings~\citep[\S6.4]{kennedy1996} run over
\emph{dimensions}, so they are exactly our coherent ones
by the following characterization (\artifact{coherent_iff_factors}):
a rescaling assigns equal factors to every same-dimension pair if and
only if it factors through a dimension rescaling. With independent base
dimensions and one base unit for each, every unit rescaling in closed
unit scope is coherent (\artifact{Scaling.coherent_of_dim_equiv}). Unit constants are excluded because $\mathbf{1}_u$ denotes the number $1$
(a value at $\Q{u}$ is a numeral at $u$, and one $u$ is the numeral $1$),
and self-relatedness would demand $1 = \psi(u)$: a term that can
\emph{name} a unit can detect a rescaling. Three term forms depend on a
unit, at three different costs: $\mathbf{1}_u$ names one and leaves the theory,
conversion reads the declared magnitudes and costs coherence, and a
comparison sees only within a single unit and costs nothing, both sides
moving by the same positive factor. Zero alone is
scale-invariant,\footnote{Whether there is one zero or many is the
subject of Russell's Chapter~XXII~\citep[\S\S172--178]{russell1903}: he
canvasses a limit construction that makes zero single and ``not one
among the magnitudes whose zero it is'' (\S176), then concludes that
each kind of magnitude has its own zero (\S177). \calc{} takes both
sides, and the seam is exactly the one between types and semantics:
$0\cdot\mathbf{1}_{\mathsf m} : \Q{\mathsf{m}}$ and $0\cdot\mathbf{1}_{\mathsf s} : \Q{\mathsf{s}}$ are distinct terms at
distinct types denoting the one real number every rescaling fixes.} the
standing \emph{zero exception}: it is why the theorem below assumes a
nonzero value. Powers need no positivity: for positive $k$,
$(k\cdot x)^{q} = k^{q}\cdot x^{q}$ at every real $x$
(\artifact{mul_rpow_of_pos_left}), so signed quantities, which Kennedy's
Coq development~\citeyearpar{kennedy2008} excludes by taking every value
positive, keep their scaling law. For negative inputs, the artifact's non-integer power operation uses
the real part of the principal complex power, rather than a real-root
convention. Theorem~\ref{thm:abs-free} certifies covariance of that
total operation;
it does not assert that such a value is a root
(Section~\ref{sec:mechanization}).

\begin{theorem}[Single-conversion invariance; \artifact{cvt_rel_iff_coherent}]
\label{thm:price-exact}
For $\cvt{x}{u}{v}$ with $x$ given any nonzero value: invariance holds
at a rescaling $\psi$ if and only if $\psi(u) = \psi(v)$, the equation
every coherent rescaling satisfies at a same-dimension pair.
\end{theorem}

Consequently, coherent rescalings are exactly those preserving every
parametric term. Sufficiency is Theorem~\ref{thm:abs-coherent}.
For necessity, a non-coherent rescaling separates some same-dimension
pair $u,v$; the parametric term $\cvt{x}{u}{v}$ at a nonzero input then
fails invariance by Theorem~\ref{thm:price-exact}. The artifact states
the biconditional for the conversions themselves: a rescaling is
coherent if and only if every single conversion of a nonzero argument is
invariant under it (\artifact{coherent_iff_cvt_invariant}).

Theorem~\ref{thm:price-exact} says the bound the abstraction theorems
place on the cost of conversion (coherence suffices) is tight, per term and per rescaling, and that
coherence is forced by the term rather than chosen by the proof. Note that multiplication cannot substitute for the
primitive, although conversion's semantics is multiplication by
$\mathrm{conv}_V(u,v)$: a literal denotes the number the programmer
wrote and $\mathbf{1}_u$ denotes the number $1$. A convert-free scalar
term denotes the same number under different valuations in independent
environments (\artifact{den_indep}), and no parametric convert-free
term converts between distinct base units
(\artifact{convert_not_definable}). The artifact runs the instance that
makes the point:
$\mathbf{1}_{\mathsf{yard}}/\mathbf{1}_{\mathsf{foot}}$ evaluates to
$1_{\mathsf{yard/foot}}$ at the declared table (\artifact{ydPerFt}), and
by \artifact{den_indep} its denotation is the same under every table.
Converted to the dimensionless unit $\mathsf{one}$ (written
$\mathsf{in}\;\mathsf{one}$), it is the declared $3$, fetched by the only
construct allowed to (\artifact{ydPerFtIn1}).
The route through feet,
$(\mathbf{1}_{\mathsf{yard}}\;\mathsf{in}\;\mathsf{foot})/\mathbf{1}_{\mathsf{foot}}$,
is the same $3$, dimensionless outright (\artifact{ydPerFtViaFt}).\footnote{$\mathsf{one}$ is the SI's name: quantities of dimension one
carry ``the unit one, symbol 1''~\longcitep{Bureau International des Poids et
Mesures}{sibrochure2019}; here it is the
empty exponent vector, whose dimension the object-oriented
design~\citep{allen2004} named \textsf{Unity}.}

\section{Accumulated Ratios, and a Decidable Diagnostic}
\label{sec:twist}

With the abstraction theorems in hand, we ask what many conversions
accumulate to, and turn the answer into a decision procedure. The
procedure is not a typing rule: every program it examines is already
well-typed and runs once its conversion factors are supplied
(Section~\ref{sec:declarations}). For a first-order program it delivers
one of three verdicts: the program is certified invariant under every
rescaling; its departure from invariance is named by one
\emph{accumulated conversion ratio}, an element of the group of units,
which we call the program's \emph{drift}; or the analysis
\emph{declines}. In declaration terms, the first verdict says that no
error in the declaration set can change the program's answer
(\artifact{den_indep_of_driftFree}), and the second says by how much one
would (\artifact{den_comp_of_drift}). For a drift-free program, a
compiler may rescale units independently; for any analyzed program, a
rescaling that sends its drift to $1$ is licensed, and at a nonzero
output no other rescaling is.

The ratio records how a value moves under a rescaling beyond the factor
its type predicts. A conversion $\cvt{e}{u}{v}$ leaves its value moving
with $u$ where its type $\Q{v}$ predicts $v$: it contributes $u/v$. A
product multiplies ratios. A sum requires its summands to agree, since
unequal excess factors do not in general give the sum a single factor;
a conditional requires the same of its two branches and of the two
quantities compared, so that a wrong declaration cannot change which
branch runs. $\log$ and $\exp$ accept drift-free arguments and decline
drifting ones, because $\log$ turns an excess factor into an excess
summand and $\exp$ into an excess exponent, and neither is a ratio. The
unit constant $\mathbf{1}_u$ is declined unconditionally:
Theorem~\ref{thm:abs-free} places it outside the invariance theory, and
declining it keeps a reported drift meaning one thing. So the
meter-to-feet round trip $(x\;\mathsf{in}\;\mathsf{ft})\;\mathsf{in}\;\mathsf{m}$
gets the trivial ratio $1$, the one-way $x\;\mathsf{in}\;\mathsf{ft}$
gets $\mathsf{m}/\mathsf{ft}$, and triviality is decidable because
ratios, like units, are exponent vectors over $\mathbb{Q}$
(\artifact{Tw.nfOne_eq_one_iff}).

\begin{theorem}[Drift decides invariance; \artifact{Twist.invariant_iff}]
\label{thm:drift}
A first-order program over scalars, with result type $\Q{u}$ and
nonzero denotation, to which the analysis assigns a ratio is invariant under \emph{all} rescalings,
that is, its denotation at $V\psi$ on inputs rescaled by $\psi$ is
$\psi(u)$ times its denotation at $V$ for every $\psi$, exactly when
that ratio evaluates to $1$ under every rescaling.
\end{theorem}

The analysis computes the ratio (\artifact{unitDrift_spec}) and decides
the condition. The sum rule is where a decline is most likely to
surprise. Both summands of $(x\;\mathsf{in}\;\mathsf{ft}) + y$, for $x$
a meter input and $y$ a foot input, have type $\Q{\mathsf{ft}}$, yet the
analysis declines the sum (\artifact{addMixed}), and rightly: at
$x = y = 1$ with $V(\mathsf m) = V(\mathsf{ft}) = 1$ the sum is $2$;
doubling the meter doubles $x$ and $V(\mathsf m)$ while $y$ and
$V(\mathsf{ft})$ stay put, and the sum becomes $2 \cdot 2/1 + 1 = 5$,
where the type $\Q{\mathsf{ft}}$ promised $2$. Over two meter inputs,
$(x\;\mathsf{in}\;\mathsf{ft}) + (y\;\mathsf{in}\;\mathsf{ft})$ is
accepted at drift $\mathsf{m}/\mathsf{ft}$ (\artifact{addTwoVars}): both
summands move together. Declines are conservative, not complete:
$0\cdot(x\;\mathsf{in}\;\mathsf{ft})+y$ is declined because its
summands' ratios disagree, yet it denotes $y$ and is invariant.
Exactness concerns assigned ratios, whose triviality is decided exactly.

Inputs are measurements and rescale with their units, so every input
receives the trivial ratio, whether it is a context variable or a
leading $\lambda$. A $\lambda$ \emph{inside} the term is different: its
argument may arrive with a drift of its own, so its bound variable
receives a \emph{ratio variable} standing for that unknown drift.
Comparison of ratios is exact, without reduction fuel, whenever the
context holds only scalar, vector, or matrix variables
(\artifact{Tw.normEq_firstOrder_iff}); internal higher-order
applications and unit binders are unrestricted. Contexts containing
unknown functions or unit-indexed families fall back to a bounded
comparison that is sound but not proved complete.

The artifact's ballistics kernel (\artifact{Ballistics}) computes a
velocity squared, an energy per unit mass, from a distance in feet and a
time in seconds, and reports it four ways: back in its input units,
certified drift-free; in metric, drift $\mathsf{ft}^2/\mathsf{m}^2$; two
routes summed, accepted at the shared drift; and through a square root,
drift $\mathsf{ft}/\mathsf{m}$, the exponent halved with the unit's. The
generic caster of Section~\ref{sec:calculus} has the drift
$\mathrm{u}/\mathrm{v}$, a ratio containing unit variables, and the
round trip through the caster, out to $\mathrm{v}$ and back to
$\mathrm{u}$, is certified drift-free \emph{without} instantiation
(\artifact{casterRound}).

Section~\ref{sec:semantics} separated the two actions of a rescaling, on
the values and on the valuation; the drift keeps them apart. With its
inputs held fixed, a program of drift $w$ is multiplied by $\varphi(w)$
when the valuation is rescaled by $\varphi$
(\artifact{den_comp_of_drift}), so a drift-free program denotes the same
number under every valuation (\artifact{den_indep_of_driftFree}), and by
adequacy (Section~\ref{sec:erasure}) so does the real-arithmetic
evaluator for a closed dimensionless one
(\artifact{evalC_indep_of_driftFree}). With the valuation held fixed
instead, a drift-free program obeys Theorem~\ref{thm:abs-free}'s scaling
law although it converts (\artifact{scaleLaw_of_driftFree}); this is the
hypothesis Section~\ref{sec:pi} uses. Both specialize one law in which
the values are rescaled by $\psi$ and the valuation by $\varphi$ at
once: the factor is $\psi(u)\cdot\varphi(w)\cdot\psi(w)$ for a program
of drift $w$ at result unit $u$ (\artifact{unitDrift_law}). The
diagnostic extends to first-order programs at vector and matrix types,
with a drift per entry (\artifact{unitDriftGen},
\artifact{scaleLaw_lin_of_driftFree_gen}): the Jacobi sweep of
Section~\ref{sec:hart} converts nowhere, and every entry receives the
trivial drift.

\section{Adequacy and Erasure}
\label{sec:erasure}

Two theorems connect the semantics to the machine. They relate
Section~\ref{sec:dynamics}'s evaluator, which carries a unit on every
value and performs the dynamic checks (we call it \emph{instrumented}),
to an \emph{erased} evaluator that carries neither; both are developed
in the artifact (modules \artifact{LambdaS.Adequacy} and
\artifact{LambdaS.Erasure}).

\begin{theorem}[Factors reach the machine;
\artifact{evalC_convert_declared}]
\label{thm:factors-reach-machine}
Let $V$ satisfy the declaration $u = q\,v$, with $u$ a base unit and $v$
a unit of its dimension, and let $e$ be a closed term of type $\Q{u}$. Instantiate the evaluator with real arithmetic and
the conversion function that $V$ determines. Then
$\cvt{e}{u}{v}$ evaluates to the denotation of $e$ multiplied by exactly
$q$, at unit $v$.
\end{theorem}

A chain thus runs from the declaration, through the semantics, to the
evaluator's multiplication. The link from semantics to evaluator is
adequacy (\artifact{evalC_eq_den}): with the conversion factors the
valuation determines, the evaluator instantiated with real arithmetic
computes the denotation. With the
yard declared at $3$ feet, the foot at $0.3048$ meters, and,
redundantly, the yard at $0.9144$ meters, one yard evaluates to $0.9144$
meters through feet or directly, and the routes are provably equal
(\artifact{one_yard_in_meters}, \artifact{one_yard_in_meters_via_feet},
\artifact{yard_routes_agree}). Separately, the compiled
\textsf{Float} binary prints $91.44$ for one hundred yards by either
route; this executed example is not a real-to-\textsf{Float} theorem.

\begin{theorem}[Erasure; \artifact{erasure_correct}]
\label{thm:erasure}
For any numeric carrier, using the same arithmetic and conversion
function in both evaluators, every closed well-typed term of scalar type
has a fuel bound at
which the instrumented evaluator returns a magnitude carrying exactly
the unit its type predicts and the erased evaluator returns exactly that
magnitude.
\end{theorem}

Units can thus be stripped from run-time values, together with every
dynamic check, without moving the numbers. Without conversion this is
Kennedy's dimension-erasure
theorem~\citeyearpar[Theorem~5.4]{kennedy1996}.
\citet[\S4.1]{kennedy2009} states the requirement: programs with and
without unit annotations behave the same. Conversion reads units, so it
is where that requirement is at risk. Behind the theorem is a
simulation (\artifact{eeval_erase}). For any term, under any number of
enclosing unit and dimension binders and any environment, if the
instrumented evaluator succeeds at fuel $n$ then the erased evaluator
succeeds at the same fuel and returns the erased value. The simulation
needs no typing hypothesis; the instrumented run's success already
witnesses that every check the erased run skips would have passed. With
adequacy, the erased evaluator instantiated with real arithmetic
computes the denotation
(\artifact{eeval_den}).
Two things deliberately survive erasure, and neither is a unit tag on a
value. Array extents survive, because a matrix with zero rows has no
entries from which to recover its width. And the unit and dimension
\emph{environments} survive, because a polymorphic conversion takes its
factor from a unit supplied at run time: data the size of the scope, not
of the payload, passed in dictionary-passing
style~\citep{wadlerblott1989}. Units are static except at the finitely
many scope entries that polymorphic conversion consults.

\section{Dimensional Analysis}
\label{sec:pi}

Two results remain: a non-definability theorem, which justifies a
constructor of \calc, and the Pi theorem of dimensional analysis, the
classical consequence of the abstraction theorem. The first is the
non-definability result of \citet{kennedy1997} in our setting. For
$u \neq 1$, no term built from variables, rational literals, and the
field operations computes square root at $\Q{u^2} \to \Q{u}$
(\artifact{sqrt_not_definable}; at the term level,
\artifact{sqrt_not_definable_tm}), so the powers of
Section~\ref{sec:calculus} must be primitive. Kennedy's proof is
semantic. Ours is linear algebra over the syntax, which the
absence of recursion permits: from an argument at $u^2$ and
dimensionless literals, the field operations reach exactly the
$\mathbb{Z}$-span of $u^2$, while $u = (u^2)^{1/2}$ needs the
$\mathbb{Q}$-span. Newton's Method, $x_{n+1} = (x_n + a/x_n)/2$, does not
escape it. Without the primitive, no term built from the field
operations manufactures the seed $x_0 : \Q{u}$ from $a : \Q{u^2}$
(\artifact{no_newton_seed_tm}); with
$\mathbf{1}_u$ as seed a finite unrolling is writable, but it denotes a
rational function of $a$, and no rational function is a square root.
The seed fixes a scale, and the iterates show it: from $\mathbf{1}_u$
on $4\,\mathsf m^2$, one step gives $2.5\,\mathsf m$ in meters and
$200.005\,\mathsf m$ in centimeters
(\artifact{LambdaS.NonDefinability}). The theorem is about exactness,
not approximation; a primitive root is what a units calculus needs to
\emph{denote} $\sqrt{\ell/g}$, and it is parametric where a seeded
iteration is not.

The second result we illustrate with a common example. What determines
the period $T$ of a pendulum? Suppose we have decided that it could
depend on the mass $m$ of the bob, the length $\ell$ of the arm, the
gravitational acceleration $g$, and the release amplitude $\theta$:
$T = f(m, \ell, g, \theta)$ for some unknown $f$. Record the exponents
of each argument's unit over the base units of mass, length, and time
as the columns of a matrix: $g$ is an acceleration,
$\mathsf{L}/\mathsf{T}^2$, and the amplitude is an angle,
dimensionless, so its column is zero. This example uses one base unit
per independent dimension, so the unit rows and the dimension rows
coincide:
\[
A \;=\;
\iflatexml
% LaTeXML, which renders arXiv's HTML, does not support \bordermatrix.
\begin{array}{c|cccc}
           & m & \ell & g  & \theta \\ \hline
\mathsf{M} & 1 & 0    & 0  & 0      \\
\mathsf{L} & 0 & 1    & 1  & 0      \\
\mathsf{T} & 0 & 0    & -2 & 0
\end{array}
\else
\bordermatrix{
           & m & \ell & g  & \theta \cr
\mathsf{M} & 1 & 0    & 0  & 0      \cr
\mathsf{L} & 0 & 1    & 1  & 0      \cr
\mathsf{T} & 0 & 0    & -2 & 0      \cr
}
\fi
\]
A power product $m^{c_1}\ell^{c_2}g^{c_3}\theta^{c_4}$ of the arguments
is dimensionless exactly when $Ac = 0$, so the dimensionless
combinations are the kernel of $A$. Here $A$ has rank $3$, the kernel has
dimension $4 - 3 = 1$, and solving $Ac = 0$ forces $c_1 = c_2 = c_3 = 0$:
the amplitude is the only dimensionless combination. The output $T$ has
exponent vector $b = (0, 0, 1)$, and $AX = b$ has the solution
$X = (0, \tfrac{1}{2}, -\tfrac{1}{2}, 0)$, the power product
$\ell^{1/2} g^{-1/2} = \sqrt{\ell/g}$. The Pi theorem says that these
two computations determine $f$ up to one unknown function of one
variable, $T = \sqrt{\ell/g} \cdot G(\theta)$, and mechanics contributes
only $G$, with $G(\theta) \to 2\pi$ at small amplitude. Note what the
mass row has already decided: mass occurs in one argument and not in
the output, so every solution and every kernel vector has zero in the
mass coordinate. The period's independence from the bob's mass, which mechanics
derives from the equation of motion, is read off the type
(\artifact{pendulum_mass_absent}).

This computation is the Pi theorem as
Buckingham~\citeyearpar{buckingham1914} stated it: a physically
meaningful equation among $n$ quantities whose units involve $r$
independent base dimensions can be rewritten as an equation among $n - r$
dimensionless products of those quantities. The kernel of $A$ supplies
the products, and there are $n - r$ of them because $A$ has rank $r$.
\citet{kennedy1997} showed that the theorem is a consequence of
parametricity: a well-typed program cannot depend on the choice of
units, and a function that cannot depend on the choice of units is a
function of dimensionless products. His calculus has no conversion. The
question for ours is whether the theorem survives it, and it does.

\begin{theorem}[Pi, for programs with conversion;
\artifact{den_pi_coherent_dichotomy}]
\label{thm:pi}
Let a parametric first-order program take $n$ scalar arguments to a
scalar result, in closed unit and dimension scope; it may convert, and
may bind units and dimensions internally. Let column $i$ of
$A_{\mathrm{dim}}$ hold the dimension exponents of argument $i$, and $b$
those of the result. Then one of two cases holds. Either some power
product of the arguments has the result's dimension, that is,
$A_{\mathrm{dim}}X = b$ is solvable, and on positive arguments the
program is that power product times a function of
$n - \mathrm{rank}\,A_{\mathrm{dim}}$ dimensionless products of its
arguments; or no power product does, and the program denotes zero on
every argument.
\end{theorem}

The pendulum is in the first case, with one dimensionless product,
$\theta$. The second case is a non-definability theorem in disguise.
Take a program that computes a period from a mass alone. Doubling the
unit of time while fixing the kilogram is a coherent rescaling. It
leaves the argument unchanged, and so the program's output, while the
scaling law demands that the output double; so the output is zero. No
power of a kilogram is a unit of time, and a program that respects rescaling
cannot manufacture one. The zero case needs no positivity of the
arguments; only the factorization does.

The theorem holds with conversion because
Theorem~\ref{thm:abs-coherent} does. A coherent rescaling preserves
every same-dimension conversion factor, and the evaluator consults only
those factors (\artifact{eval_congr_sameDim}), so the program computes
the same number at $V$ and at $V\psi$; by normalization and adequacy its
denotation obeys the dimension-level scaling law
(\artifact{den_mulScaleLaw_coherent}), even where the drift analysis
declines it (\artifact{addMixed_coherent} is the instance for
Section~\ref{sec:twist}'s declined sum). The scope is closed because
the proof runs through the evaluator, whose termination theorem is for
closed terms; a direct semantic proof would lift the restriction. From
the scaling law the two cases are linear algebra
(\artifact{mulScaleLaw_dichotomy}): the $n - r$ coordinates
$\sum_i c^j_i \log x_i$, for a rational basis $c^1, \ldots, c^{n-r}$ of
the kernel, are constant on rescaling orbits and separate them, so a
function invariant under rescaling is a function of them
(\artifact{invariant_descends},
\artifact{mulScaleLaw_factorization_reduced}), and with $\mathbb{Q}$
exponents rank-nullity gives the count (\artifact{pi_count}) where
Kennedy's $\mathbb{Z}$-exponent proof needs Smith normal form.
Every step is mechanized, with the equivalence between functions
obeying a scaling law and scale-invariant functions proved in both
directions (\artifact{piEquiv}, \artifact{piEquivSigned});
Section~\ref{sec:related} compares the statement with Kennedy's.

A stronger law is available to programs the drift analysis certifies.
Independent unit rescalings, which move the meter without the foot,
form a larger group than dimension rescalings, and a drift-free
program respects all of them (\artifact{den_mulScaleLaw_driftFree}), as
does every parametric convert-free one (\artifact{den_mulScaleLaw});
neither needs the closed scope the coherent version does.
For these programs the matrix may have a row for each base unit and
each unit variable rather than for each dimension, and the two cases
hold over that matrix too. The constraint is stronger: a meter argument
divided by a foot argument is dimensionless but has unit
$\mathsf m/\mathsf{ft}$, so the unit-level theorem does not let the
program depend on it, where the dimension-level theorem does; and a
program from a meter argument to a foot result, dimensionally
consistent as it is, falls in the zero case, since no power of a meter
is a foot.

\section{Dimensioned Linear Algebra}
\label{sec:hart}

Dimensioned linear algebra is where units matter most in numerical
code, and it is where the invariance theory of
Sections~\ref{sec:semantics}--\ref{sec:twist} meets code a numerical
analyst would write. This section shows that the type discipline those
theorems assume for vectors and linear maps is \citeauthor{hart1995}'s
\citeyearpar{hart1995}, that it costs nothing at run time, and that on a
real kernel the theorems say two things untyped code cannot.

A vector of dimensioned entries poses no problem: each component
carries its own unit, and a state of position and momentum lives in the
space $[\mathsf{m},\ \mathsf{kg}\cdot\mathsf{m}/\mathsf{s}]$. A matrix
does. Hart observed that a matrix of dimensioned entries cannot carry
arbitrary units: for $x$ in $\mathsf{Vec}\,\vec{u}$ and $Ax$ in
$\mathsf{Vec}\,\vec{w}$, each summand $A_{ji}\,x_i$ of
$(Ax)_j = \sum_i A_{ji}\,x_i$ must carry the unit $w_j$ of the result
component, so entry $(j,i)$ carries $w_j/u_i$, and the whole array of
entry units is determined by two vectors of units, one for the domain
and one for the codomain. We say the array of entry units has
\emph{rank one}; this says nothing about the numerical rank of the
matrix. In \calc, as in the type systems of
\citet{griffioen2015,griffioen2019} and of \citet{mcbride2022}, the
invariant is representational: a type $\mathsf{Lin}\,\vec{u}\,\vec{w}$
carries its two spaces and nothing per entry, and the introduction rule
\ftrule{T-MCons} (Figure~\ref{fig:typing}) accepts a row only at the
space $w/\vec{u}$, so no term constructs a matrix outside Hart's form.
Type soundness carries the two spaces, over an array of bare
magnitudes, to the evaluator (\artifact{lin_soundness_total}); since
the two spaces determine every entry unit, the array loses nothing, and
the compiled evaluator passes vectors and matrices unboxed to a
BLAS-shaped kernel~\citep{lawson1979} (Section~\ref{sec:mechanization}).
The eliminator $\mathsf{row}_j$ returns a row at the type
\ftrule{T-MCons} demanded, so entries and transposes are writable
(\artifact{fromTimeT}).

Hart went on to classify dimensioned matrices by the operations their
units permit, and in \calc{} each class is an identity of unit
assignments, proved in a model of dimensioned matrices (module
\artifact{LambdaS.Map}) in which the unit of entry $(j,i)$ is
\emph{defined} as $w_j/u_i$ (\artifact{entry}), the unit
\ftrule{T-MCons} assigns (\artifact{HasTy.mcons_entry}). A composite's
summand $A_{kj}B_{ji}$ carries $(w_k/v_j)(v_j/u_i) = w_k/u_i$, so
Hart's ``multipliable'' condition is not a condition to check but the
only composition writable (\artifact{entry_comp}). On an
\emph{endomorphism} $\mathsf{Lin}\,\vec{u}\,\vec{u}$ every diagonal
entry carries $u_i/u_i = 1$, and so does every product
$\prod_i A_{\sigma(i)\,i}$ over a permutation $\sigma$, so trace and
determinant are plain numbers (\artifact{entry_id_diag},
\artifact{entry_perm_prod}): on the position-momentum space the entry
units are
$\left(\begin{smallmatrix}1 & \mathsf{s}/\mathsf{kg}\\
\mathsf{kg}/\mathsf{s} & 1\end{smallmatrix}\right)$, and written as
terms the trace and determinant typecheck at unit $1$ and the
$2\times 2$ cofactor inverse as an endomorphism of the same space
(\artifact{traceTm}, \artifact{detTm}, \artifact{invTm}), with the
compiled binary checking that $A^{-1} \circ A$ evaluates to the
identity. A space is \emph{uniform} when every component carries one
unit, as $[\mathsf m, \mathsf m]$ does; between uniform spaces every
entry carries the same unit (\artifact{svd_entry_const}), which is what
sorting singular values requires, and a space equals its \emph{dual},
the space of componentwise reciprocal units, exactly when it is
dimensionless (\artifact{transpose_comp_direct_iff}), which is what
composing a Cholesky factor with its transpose requires
(\artifact{cholesky_factor_dimensionless}). These are identities of
unit assignments; they do not implement a decomposition. The kernel
below supplies a concrete typing test.

A $2\times 2$ Jacobi step (\artifact{sweep}) rotates a symmetric matrix
toward diagonal form to compute its eigenvalues. A symmetric matrix is
the matrix of a quadratic form, taking a vector $x$ to the covector
$y \mapsto y^{\mathsf T} A\,x$, so on the uniform space
$U_2=[\mathsf m,\mathsf m]$ its type is $\mathsf{Lin}\,U_2\,U_2^{-1}$
with $U_2^{-1}=[1/\mathsf m,1/\mathsf m]$ its dual, and every entry
carries $\mathsf m^{-2}$. The step computes
$\tau=(A_{22}-A_{11})/(2A_{12})$, chooses the corresponding cosine $c$
and sine $s$, and forms $R^{\mathsf T}AR$ with
$R=\left(\begin{smallmatrix}c&s\\-s&c\end{smallmatrix}\right)$, which
zeroes the off-diagonal pair in exact arithmetic; the compiled
numerical checks exercise both signs of the off-diagonal and equal,
unequal, and already-diagonal inputs within a floating-point tolerance.
Two things about this kernel are invisible in untyped code.

First, $\tau$ is a ratio of entries at one unit and so is
dimensionless, and cosine and sine follow: the rotation is a matrix of
plain numbers whatever the space's unit. On a \emph{non-uniform} space
the kernel does not typecheck at all: its two diagonal entries carry
different units, so the numerator of $\tau$ is an addition at unequal
units ($\Gamma_N$ in \textsf{Examples.lean}). Hart's uniformity
condition is not a hypothesis a programmer must carry; it is the
condition under which the code can be written.

Second, a stopping test compares $|A_{12}|$ against a tolerance
carrying the entries' unit, and the invariance theory distinguishes
how that tolerance is supplied. A \emph{relative} test,
$\varepsilon\max(|A_{11}|,|A_{22}|)$ for dimensionless $\varepsilon$,
scales a quantity already in hand, names no unit constant, and stays
inside the parametric fragment, so Theorem~\ref{thm:abs-free} applies
to the kernel. A fixed absolute tolerance written as a literal times
$\mathbf{1}_u$ names a unit and leaves that fragment
(\artifact{stopAbsolute}); both forms typecheck, and both run. An
absolute tolerance supplied as an input rescales with its unit and
remains parametric, so what the theory forbids is not absolute
tolerances but tolerances that fix a scale in the program text. The
drift analysis of Section~\ref{sec:twist} certifies the sweep itself
drift-free, entry by entry. The kernel examples test one stopping
decision; iteration is outside a calculus without recursion.

\section{Mechanization}
\label{sec:mechanization}

The modules cited throughout develop each result in full, with rendered
documentation. The development is $11{,}100$ lines of Lean~4~\citep{lean4} definitions and
proofs and $6{,}600$ lines of documentation (the artifact's
\textsf{count\_lines} script counts \textsf{LambdaS/*.lean}), building with no admitted
proofs (Lean's \textsf{sorry}) and no axioms beyond the three of Lean's
standard library.\footnote{Propositional extensionality, choice, and quotient
soundness. The kernel, Lean's proof checker, checks every proof as part
of the build; that no theorem our paper cites depends on a further axiom
is audited by a script in the artifact's continuous integration.} It stays that small because units are
exponent vectors (Section~\ref{sec:calculus}), as in Kennedy's Coq
development~\citeyearpar{kennedy2008} (over $\mathbb{Z}$ there): substitution is a linear map, and where Kennedy's
substitution lemma, over types interpreted as Coq types, needs equality
coercions he calls awkward, ours keeps types syntactic and reorders a
finite sum.

Two things beyond the theorems. First, a defect the
mechanization caught that every test missed: a variable-capture bug in
substitution, invisible to the weaken-then-substitute tests, where it
canceled itself, exposed only when the composition law of
substitutions was proved. The artifact represents variables as de Bruijn
indices~\citep{debruijn1972}, with types and terms indexed by the number
of enclosing unit and dimension binders; the named variables of Figures~\ref{fig:syntax}
and~\ref{fig:typing} are a presentation choice. Scope indices rule out
dangling unit and dimension variables, but not selection of the wrong
variable within a scope. The defect illustrates why substitution needs identity and
composition laws in addition to examples.

Second, the trusted base. A reader who believes a
theorem trusts Lean's kernel and its three axioms. Concrete checker
and diagnostic outcomes in \textsf{Examples.lean}, including the caster,
Jacobi sweep, its non-uniform rejection, and the reported drift tables,
are mostly \textsf{\#guard} assertions, executed by Lean's compiler
rather than checked by its kernel.
Their execution adds the compiler to the trusted base. Numerical
regression checks, including the Jacobi residual and stopping decisions,
run in the native binary with \textsf{Float}. Kernel-checked typing derivations and
parametricity proofs are separate from these executed checks.
A reader who
believes the number the compiled binary prints trusts, in addition,
Lean's code generator, the \textsf{Float} carrier, and three C
functions behind a BLAS-shaped interface: a dot product, a
matrix-vector product, and a query naming the backend. The denotational theorems use real arithmetic; they do not prove
that floating-point evaluation equals it. Rounding affects defined
operations. The carriers also totalize partial operations differently:
real division by zero and $\log 0$ yield $0$, while the binary follows
its floating-point operations. Non-integer powers of negative values
likewise use different conventions (Section~\ref{sec:semantics}); no
real-to-floating-point correspondence is claimed at these points. We
claim no IEEE conformance; the standard~\citep{ieee754-2019} itself
requires correctly rounded field operations and square root, while the
transcendental functions are recommended operations, correctly rounded
only when provided.

\section{Related Work}
\label{sec:related}

In this section, we situate \calc{} among the calculi it descends from, the
practical systems that motivated it, and the proof methods it borrows.

\paragraph{Kennedy.}
\citet{kennedy1997} is the direct ancestor: a typed lambda calculus with
units as a free abelian group, unit polymorphism, the abstraction theorem,
non-definability of square root, and the Pi theorem as a type
isomorphism derived from parametricity.
Kennedy established several surprising properties of unit systems, and
nearly every design decision in \calc{} inherits from it or answers it.
The differences are the ones our paper is about: Kennedy's
calculus has no conversion operator (nothing in it can observe a unit),
no dimensions distinct from units, and $\mathbb{Z}$ rather than
$\mathbb{Q}$ exponents. The later mechanization~\citep{kennedy2008}
interprets types shallowly into Coq (``units ignored'') over a base
domain restricted to positive values. It has no terms yet, only
semantic results over total Coq functions, so neither the bottom element
that domain theory forced nor the question of totality arises; his 2008
work-in-progress list includes formalizing the Pi theorem and
non-definability. His Pi theorem (Theorem~3) states the solvable case
for $n$ arguments over $\mathbb{Z}$ exponents, by Smith normal form, as
a type isomorphism for positive values, and his Section~5.2 shows every
term of unsolvable signature trivial. Ours (Section~\ref{sec:pi})
states both cases as one dichotomy with outputs of either sign, has
$\mathbb{Q}$ exponents, covers base units and unit variables jointly,
and mechanizes the equivalence with both round trips proved.

\paragraph{Atkey, Johann, and Kennedy.}
\citet{atkey2013} generalize the theory to types indexed by arbitrary
algebraic structures, with an abstraction theorem over arbitrary index
theories and a general non-definability criterion for first-order types
that subsumes Kennedy's square-root result and ours. Our development is
narrower and deeper on one axis: in their units
instance the index is a group and no term reads it, which is why their
abstraction theorem is unconditional on the term language, where \calc{}
has an operator that does read it. In that configuration coherence becomes
a theorem rather than a
vacuous condition. Their scaling law is a type isomorphism that splits off
the zero argument as one factor of a product, the same exception our
nonzero hypotheses mark. From it their paper derives a one-variable
instance of the Pi theorem,
$\forall s.\,\mathsf{real}\langle s\rangle \to \mathsf{real}\langle s^2\rangle
\cong \mathsf{real}\langle 1\rangle$, the area of a circle. Their Coq
development contains the scalings, non-inhabitation, and the square and
cube root results, and proves the zero-argument isomorphism (the only
polymorphic scalar is zero); the instance
$\forall s.\,\mathsf{real}\langle s\rangle \to \mathsf{real}\langle s\rangle
\cong \mathsf{real}\langle 1\rangle$ is stated there in a comment with
its proof unfinished.

Dimensions as a grading and several units per dimension are
already \citeauthor{mcbride2022}'s \citeyearpar{mcbride2022}: their semiring is graded over the group of
dimensions, so that a length in feet and a length in meters already
inhabit one graded component, and a unit system is a choice of one
unit per dimension used to render a quantity numerically.
What is new here is a term former that reads the choice, and the coherence
condition reading it forces.

\paragraph{Practical systems.}
\citet{allen2004} formulate dimensions and units as classes with statically
typed metaclasses (given a core calculus by
\citet{tobinhochstadt2005}), dynamic conversion factors, nonlinear scales,
and
several units per dimension: the practical feature set \calc{} gives a
theory to, minus the scales. The design entered the Fortress
specification~\citep[\S2.10]{fortress2006}, whose \textsf{in} converts a
measurement to another unit of the same dimension: \calc{} takes both its
notation and its side condition from that operator. The core calculus
covered the metaclass mechanism rather than conversion, and no invariance
theorem was proved for either. F\#'s units~\citep{kennedy2009}
erase completely and convert by multiplication the programmer writes
herself; the type system checks the unit annotation on
the literal factor but nothing ties the number $3.28084$ to feet and meters,
and Theorem~\ref{thm:factors-reach-machine} is that tie.
\citet{gundry2015} implements unit inference in GHC with a typechecker
plugin solving free-abelian-group constraints. That design indexes by
units alone and leaves separate dimensions and inferred conversions as
further work. Our rational elimination lemmas address unit equations;
they do not settle the two-level inference question of
Section~\ref{sec:calculus}.
Osprey~\citep{jiangsu2006} checks the factors C programmers write
against magnitudes carried in its units, and caught a mile-to-meter
factor typed as $1682$ in an annotated test program; it has no
invariance theory.
\citet{karrloveman1978} proposed declared factors checked by row
reduction of their logarithms, over rational exponents, with the
compiler inserting the conversion factor the declarations determine;
the Pascal proposals that followed reached the same
problem from the language side, and \citet{house1983} was alone among
them in seeing that units need
polymorphism~\citep[\S1.3]{kennedy1996}. \citet{tranconlepper2025} give declared conversions
an algebraic theory. Their Theorem~4.38 establishes consistency for
well-defining conversions, and Remark~4.40 conjectures residual
finiteness as a route to deciding convertibility of closed conversions.
In our rational-exponent model, Theorem~\ref{thm:consistency}
characterizes consistency of finite declarations with positive rational
factors; the verified checker decides this without an acyclicity
hypothesis.
\citet{fosterwolff2023} build the ISQ (the
International System of Quantities) in
Isabelle/HOL with the same exponent-vector representation of dimensions we
use, $\mathbb{Z}$ exponents, and conversion between unit \emph{systems}.
The unit system, not the unit, is their type-level grading: yard and
foot differ only at the value level, a distinction our types draw and
theirs do not. Their formalization has no language calculus and no
invariance theory; it is a
metrology library with proof automation, and a good one.
\citet{bottajansson2026} formalize unit systems, the covariance
principle, and the statement of the Pi theorem in Idris, with
$\mathbb{Z}$ exponents, and observe that its proofs, being
non-constructive, are not implementable there, the obstacle being the
excluded middle. Ours use classical logic, as Lean's real analysis does; they
are proofs, not programs that compute the factorization.
\citet{danish2024} report on incremental units verification for Fortran
at scale, in a preprint; their checker found the wrong conversion factor
cited in our introduction. \citet{haynes2026} combines Kennedy-style dimensional inference
with numeric representation and memory placement in Clef.
His account does not develop conversion semantics or a coherence theorem;
\calc{} makes conversion primitive and characterizes the rescalings that
preserve every parametric term, while leaving principal-type inference
with conversion open.

\paragraph{Methods.}
Fuel-based definitional interpreters and soundness-as-preservation over
them are standard~\citep{reynolds1972,aminrompf2017,owens2016}, as are
intrinsically typed interpreters~\citep{poulsen2018} and the discharge of
the fuel by a normalization theorem; we follow all three.
The accumulated-ratio analysis of
Section~\ref{sec:twist} is in the family of graded and coeffect
systems~\citep{petricek2014,gaboardi2016,orchard2019}: each judgment
carries an element of an algebra recording how a term uses its context,
here the net conversion. We know of no prior grading by conversion. The
nearest operation is the constant-multiplication primitive of
\citeauthor{atkey2013}'s \citeyearpar{atkey2013} distance-indexed instance (their \S7), which records
the factor in its index; it is assumed in an operation context over a
distance bound, not written as a term former over a dimension grading.
Among language operators, the nearest is Fortress's \textsf{in}, which no
calculus has graded.
Per-component units for linear algebra are due to \citet{hart1995};
\citet{griffioen2015,griffioen2019} types them with inference, and
\citet{mcbride2022} with dependent types graded by dimension. Typed index
sets for array programs, as in Dex~\citep{paszke2021}, are the natural
neighbors of our $\mathsf{Vec}$/$\mathsf{Lin}$ types.

\section{Conclusion}
\label{sec:conclusion}

We have presented a calculus in which conversion coexists with the theory
that makes unit types worth having. Every parametric term is invariant under coherent rescalings.
For a first-order program with nonzero output to which the analysis
assigns a ratio, invariance under all rescalings is equivalent to a
trivial ratio; the ratio records dependence on declared
magnitudes, and triviality is decidable. The treatment extends to
dimensioned vectors and linear maps (Section~\ref{sec:hart}), and the
declared factors reach the real-arithmetic evaluator's multiplication by theorem.
Executable declaration checking decides consistency and whether every
same-dimension factor is determined, and extracts those factors exactly.
The artifact is complete from the
checker to the compiled binary, and small enough to read, because units are
exponent vectors and their metatheory is linear algebra.

Many extensions would be necessary for \calc{} to become a usable language
in practice. A first tier adds let, fixpoints, a boolean type and its
conditional, and
size polymorphism for the vector types. Their typing rules can retain
the same unit equalities; extending the metatheory requires separate
proofs. Fixpoints would remove the normalization guarantee of
Section~\ref{sec:dynamics}, and extending the diagnostic to recursive
calls requires an account of equations for their accumulated ratios. One of the most important
extensions is a module system in which unit
declarations are interfaces, so that consistency and the availability of
required conversion factors (Section~\ref{sec:declarations}), already
checked for a finite unit system, become linking-time obligations.

A second line is completeness of the diagnostic under binders.
Comparison is exact for ratios with scalar, vector, or matrix
inputs, including internal higher-order applications, without a fuel
bound. Unknown function and unit-family inputs retain a conservative
comparison; proving sufficient reduction fuel or replacing that fallback
with a terminating normalizer remains open. Separately, distinct
argument drifts are independent unless a call site establishes their
agreement. Exploiting that agreement needs ratio families defined only
where their argument drifts agree.

Finally, scales and defined zeros~\citep{allen2004} require structure
beyond the multiplicative theory: affine offsets for temperature scales
and timestamps, and logarithmic coordinates for decibel levels.
Extending the declaration discipline, coherence theorem, and drift
analysis to these cases remains open.

\begin{acks}
This paper and its artifact were developed in sustained collaboration with
Claude Fable~5 (Anthropic). Claude drafted and revised the prose under the
author's direction, engineered much of the Lean mechanization, and
strengthened the artifact each time review exposed a gap between what the
text claimed and what the proofs delivered; the author set the design,
questioned the claims, and made the calls, including the ones Claude got
wrong first. Codex Astra (OpenAI) subsequently revised the paper in response to
human and AI reviews and extended the Pi mechanization.
Authorship policies rightly reserve authorship for those who
can be accountable for the work; the author reviewed every theorem and
every sentence and bears sole responsibility. The author thanks Scott
Kilpatrick for comments on an earlier version of this paper.
\end{acks}

\section*{Disclaimer}

This document is being distributed for informational and educational
purposes only and is not an offer to sell or the solicitation of an offer
to buy any securities or other instruments. The information contained
herein is not intended to provide, and should not be relied upon for,
investment advice. The views expressed herein are not necessarily the
views of Two Sigma Investments, LP or any of its affiliates
(collectively, ``Two Sigma''). Such views reflect the assumptions of the
author(s) of the document and are subject to change without notice. The
document may employ data derived from third-party sources. No
representation is made by Two Sigma as to the accuracy of such
information and the use of such information in no way implies an
endorsement of the source of such information or its validity.

The copyrights and/or trademarks in some of the images, logos or other
material used herein may be owned by entities other than Two Sigma. If
so, such copyrights and/or trademarks are most likely owned by the entity
that created the material and are used purely for identification and
comment as fair use under international copyright and/or trademark laws.
Use of such image, copyright or trademark does not imply any association
with such organization (or endorsement of such organization) by Two
Sigma, nor vice versa.

\bibliographystyle{ACM-Reference-Format}
\bibliography{refs}


\begin{thebibliography}{46}

%%% the Web address.

\ifx \showCODEN    \undefined \def \showCODEN     #1{\unskip}     \fi
\ifx \showDOI      \undefined \def \showDOI       #1{#1}\fi
\ifx \showISBNx    \undefined \def \showISBNx     #1{\unskip}     \fi
\ifx \showISBNxiii \undefined \def \showISBNxiii  #1{\unskip}     \fi
\ifx \showISSN     \undefined \def \showISSN      #1{\unskip}     \fi
\ifx \showLCCN     \undefined \def \showLCCN      #1{\unskip}     \fi
\ifx \shownote     \undefined \def \shownote      #1{#1}          \fi
\ifx \showarticletitle \undefined \def \showarticletitle #1{#1}   \fi
\ifx \showURL      \undefined \def \showURL       {\relax}        \fi
% invisible to TeX
\providecommand\bibfield[2]{#2}
\providecommand\bibinfo[2]{#2}
\providecommand\natexlab[1]{#1}
\providecommand\showeprint[2][]{arXiv:#2}

\bibitem[Allen et~al\mbox{.}(2006)]%
        {fortress2006}
\bibfield{author}{\bibinfo{person}{Eric Allen}, \bibinfo{person}{David Chase},
  \bibinfo{person}{Joe Hallett}, \bibinfo{person}{Victor Luchangco},
  \bibinfo{person}{Jan-Willem Maessen}, \bibinfo{person}{Sukyoung Ryu},
  \bibinfo{person}{Guy~L. {Steele Jr.}}, {and} \bibinfo{person}{Sam
  Tobin-Hochstadt}.} \bibinfo{year}{2006}\natexlab{}.
\newblock \bibinfo{booktitle}{\emph{The {Fortress} Language Specification,
  Version 1.0$\alpha$}}.
\newblock \bibinfo{type}{{T}echnical {R}eport}. \bibinfo{institution}{Sun
  Microsystems}.
\newblock


\bibitem[Allen et~al\mbox{.}(2004)]%
        {allen2004}
\bibfield{author}{\bibinfo{person}{Eric Allen}, \bibinfo{person}{David Chase},
  \bibinfo{person}{Victor Luchangco}, \bibinfo{person}{Jan-Willem Maessen},
  {and} \bibinfo{person}{Guy~L. {Steele Jr.}}} \bibinfo{year}{2004}\natexlab{}.
\newblock \showarticletitle{Object-Oriented Units of Measurement}. In
  \bibinfo{booktitle}{\emph{Proceedings of the 19th {ACM} {SIGPLAN} Conference
  on Object-Oriented Programming, Systems, Languages, and Applications
  ({OOPSLA} '04)}}. \bibinfo{publisher}{ACM}, \bibinfo{pages}{384--403}.
\newblock
\urldef\tempurl%
\url{https://doi.org/10.1145/1028976.1029008}
\showDOI{\tempurl}


\bibitem[Amin and Rompf(2017)]%
        {aminrompf2017}
\bibfield{author}{\bibinfo{person}{Nada Amin} {and} \bibinfo{person}{Tiark
  Rompf}.} \bibinfo{year}{2017}\natexlab{}.
\newblock \showarticletitle{Type Soundness Proofs with Definitional
  Interpreters}. In \bibinfo{booktitle}{\emph{Proceedings of the 44th {ACM}
  {SIGPLAN} Symposium on Principles of Programming Languages ({POPL} '17)}}.
  \bibinfo{publisher}{ACM}, \bibinfo{pages}{666--679}.
\newblock
\urldef\tempurl%
\url{https://doi.org/10.1145/3009837.3009866}
\showDOI{\tempurl}


\bibitem[Astin and Karo(1959)]%
        {astinkaro1959}
\bibfield{author}{\bibinfo{person}{Allen~V. Astin} {and}
  \bibinfo{person}{H.~Arnold Karo}.} \bibinfo{year}{1959}\natexlab{}.
\newblock \showarticletitle{Refinement of Values for the Yard and the Pound}.
\newblock \bibinfo{journal}{\emph{Federal Register}} \bibinfo{volume}{24},
  \bibinfo{number}{128} (\bibinfo{year}{1959}), \bibinfo{pages}{5348}.
\newblock
\newblock
\shownote{FR Doc. 59-5442, July 1, 1959; approved June 25, 1959 by F. H.
  Mueller, Secretary of Commerce}.


\bibitem[Atkey et~al\mbox{.}(2013)]%
        {atkey2013}
\bibfield{author}{\bibinfo{person}{Robert Atkey}, \bibinfo{person}{Patricia
  Johann}, {and} \bibinfo{person}{Andrew~J. Kennedy}.}
  \bibinfo{year}{2013}\natexlab{}.
\newblock \showarticletitle{Abstraction and Invariance for Algebraically
  Indexed Types}. In \bibinfo{booktitle}{\emph{Proceedings of the 40th {ACM}
  {SIGPLAN-SIGACT} Symposium on Principles of Programming Languages ({POPL}
  '13)}}. \bibinfo{publisher}{ACM}, \bibinfo{pages}{87--100}.
\newblock
\urldef\tempurl%
\url{https://doi.org/10.1145/2429069.2429082}
\showDOI{\tempurl}


\bibitem[Botta and Jansson(2026)]%
        {bottajansson2026}
\bibfield{author}{\bibinfo{person}{Nicola Botta} {and} \bibinfo{person}{Patrik
  Jansson}.} \bibinfo{year}{2026}\natexlab{}.
\newblock \showarticletitle{Types, Equations, Dimensions and the {Pi} Theorem}.
\newblock \bibinfo{journal}{\emph{Journal of Functional Programming}}
  \bibinfo{volume}{36}, Article \bibinfo{articleno}{3} (\bibinfo{year}{2026}),
  \bibinfo{numpages}{43}~pages.
\newblock
\urldef\tempurl%
\url{https://doi.org/10.46298/jfp.17762}
\showDOI{\tempurl}


\bibitem[Buckingham(1914)]%
        {buckingham1914}
\bibfield{author}{\bibinfo{person}{Edgar Buckingham}.}
  \bibinfo{year}{1914}\natexlab{}.
\newblock \showarticletitle{On Physically Similar Systems; Illustrations of the
  Use of Dimensional Equations}.
\newblock \bibinfo{journal}{\emph{Physical Review}} \bibinfo{volume}{4},
  \bibinfo{number}{4} (\bibinfo{year}{1914}), \bibinfo{pages}{345--376}.
\newblock
\urldef\tempurl%
\url{https://doi.org/10.1103/PhysRev.4.345}
\showDOI{\tempurl}


\bibitem[{Bureau International des Poids et Mesures}(2019)]%
        {sibrochure2019}
\bibfield{author}{\bibinfo{person}{{Bureau International des Poids et
  Mesures}}.} \bibinfo{year}{2019}\natexlab{}.
\newblock \bibinfo{title}{The International System of Units ({SI})}.
\newblock \bibinfo{howpublished}{9th edition}.
\newblock
\urldef\tempurl%
\url{https://doi.org/10.59161/AUEZ1291}
\showDOI{\tempurl}


\bibitem[Chlipala(2013)]%
        {cpdt}
\bibfield{author}{\bibinfo{person}{Adam Chlipala}.}
  \bibinfo{year}{2013}\natexlab{}.
\newblock \bibinfo{booktitle}{\emph{Certified Programming with Dependent Types:
  A Pragmatic Introduction to the {Coq} Proof Assistant}}.
\newblock \bibinfo{publisher}{MIT Press}.
\newblock
\urldef\tempurl%
\url{https://doi.org/10.7551/mitpress/9153.001.0001}
\showDOI{\tempurl}


\bibitem[Danish et~al\mbox{.}(2024)]%
        {danish2024}
\bibfield{author}{\bibinfo{person}{Matthew Danish}, \bibinfo{person}{Dominic
  Orchard}, {and} \bibinfo{person}{Andrew Rice}.}
  \bibinfo{year}{2024}\natexlab{}.
\newblock \bibinfo{title}{Incremental Units-of-Measure Verification}.
\newblock \bibinfo{howpublished}{arXiv:2406.02174}.
\newblock
\newblock
\shownote{Written in 2018 and archived in 2024}.


\bibitem[de~Bruijn(1972)]%
        {debruijn1972}
\bibfield{author}{\bibinfo{person}{Nicolaas~G. de Bruijn}.}
  \bibinfo{year}{1972}\natexlab{}.
\newblock \showarticletitle{Lambda Calculus Notation with Nameless Dummies, a
  Tool for Automatic Formula Manipulation, with Application to the
  {Church-Rosser} Theorem}.
\newblock \bibinfo{journal}{\emph{Indagationes Mathematicae (Proceedings)}}
  \bibinfo{volume}{75}, \bibinfo{number}{5} (\bibinfo{year}{1972}),
  \bibinfo{pages}{381--392}.
\newblock
\urldef\tempurl%
\url{https://doi.org/10.1016/1385-7258(72)90034-0}
\showDOI{\tempurl}


\bibitem[de~Moura and Ullrich(2021)]%
        {lean4}
\bibfield{author}{\bibinfo{person}{Leonardo de Moura} {and}
  \bibinfo{person}{Sebastian Ullrich}.} \bibinfo{year}{2021}\natexlab{}.
\newblock \showarticletitle{The {Lean} 4 Theorem Prover and Programming
  Language}. In \bibinfo{booktitle}{\emph{Automated Deduction ({CADE} 28)}}
  \emph{(\bibinfo{series}{LNCS}, Vol.~\bibinfo{volume}{12699})}.
  \bibinfo{publisher}{Springer}, \bibinfo{pages}{625--635}.
\newblock
\urldef\tempurl%
\url{https://doi.org/10.1007/978-3-030-79876-5_37}
\showDOI{\tempurl}


\bibitem[Foster and Wolff(2023)]%
        {fosterwolff2023}
\bibfield{author}{\bibinfo{person}{Simon Foster} {and}
  \bibinfo{person}{Burkhart Wolff}.} \bibinfo{year}{2023}\natexlab{}.
\newblock \showarticletitle{Automated Reasoning for Physical Quantities, Units,
  and Measurements in {Isabelle/HOL}}. In \bibinfo{booktitle}{\emph{27th
  International Conference on Engineering of Complex Computer Systems ({ICECCS}
  2023)}}. \bibinfo{publisher}{IEEE}, \bibinfo{pages}{136--141}.
\newblock
\urldef\tempurl%
\url{https://doi.org/10.1109/ICECCS59891.2023.00025}
\showDOI{\tempurl}


\bibitem[Gaboardi et~al\mbox{.}(2016)]%
        {gaboardi2016}
\bibfield{author}{\bibinfo{person}{Marco Gaboardi}, \bibinfo{person}{{Shin-ya}
  Katsumata}, \bibinfo{person}{Dominic Orchard}, \bibinfo{person}{Flavien
  Breuvart}, {and} \bibinfo{person}{Tarmo Uustalu}.}
  \bibinfo{year}{2016}\natexlab{}.
\newblock \showarticletitle{Combining Effects and Coeffects via Grading}. In
  \bibinfo{booktitle}{\emph{Proceedings of the 21st {ACM} {SIGPLAN}
  International Conference on Functional Programming ({ICFP} '16)}}.
  \bibinfo{publisher}{ACM}, \bibinfo{pages}{476--489}.
\newblock
\urldef\tempurl%
\url{https://doi.org/10.1145/2951913.2951939}
\showDOI{\tempurl}


\bibitem[Girard(1972)]%
        {girard1972}
\bibfield{author}{\bibinfo{person}{Jean-Yves Girard}.}
  \bibinfo{year}{1972}\natexlab{}.
\newblock \emph{\bibinfo{title}{Interpr{\'e}tation fonctionnelle et
  {\'e}limination des coupures de l'arithm{\'e}tique d'ordre sup{\'e}rieur}}.
\newblock Th{\`e}se d'{\'E}tat. \bibinfo{school}{Universit{\'e} Paris VII}.
\newblock


\bibitem[Goubault(1994)]%
        {goubault1994}
\bibfield{author}{\bibinfo{person}{Jean Goubault}.}
  \bibinfo{year}{1994}\natexlab{}.
\newblock \showarticletitle{Inf\'erence d'unit\'es physiques en {ML}}. In
  \bibinfo{booktitle}{\emph{Journ\'ees Francophones des Langages Applicatifs}}
  \emph{(\bibinfo{series}{Collection didactique})},
  \bibfield{editor}{\bibinfo{person}{P.~Cointe}, \bibinfo{person}{C.~Queinnec},
  {and} \bibinfo{person}{B.~Serpette}} (Eds.). \bibinfo{publisher}{INRIA},
  \bibinfo{address}{Noirmoutier}, \bibinfo{pages}{3--20}.
\newblock


\bibitem[Griffioen(2015)]%
        {griffioen2015}
\bibfield{author}{\bibinfo{person}{P.~R. Griffioen}.}
  \bibinfo{year}{2015}\natexlab{}.
\newblock \showarticletitle{Type Inference for Array Programming with
  Dimensioned Vector Spaces}. In \bibinfo{booktitle}{\emph{Proceedings of the
  27th Symposium on the Implementation and Application of Functional
  Programming Languages ({IFL} '15)}}. \bibinfo{publisher}{ACM},
  \bibinfo{pages}{4:1--4:12}.
\newblock
\urldef\tempurl%
\url{https://doi.org/10.1145/2897336.2897341}
\showDOI{\tempurl}


\bibitem[Griffioen(2019)]%
        {griffioen2019}
\bibfield{author}{\bibinfo{person}{P.~R. Griffioen}.}
  \bibinfo{year}{2019}\natexlab{}.
\newblock \emph{\bibinfo{title}{A Unit-Aware Matrix Language and its
  Application in Control and Auditing}}.
\newblock \bibinfo{thesistype}{Ph.\,D. Dissertation}.
  \bibinfo{school}{Universiteit van Amsterdam}.
\newblock
\urldef\tempurl%
\url{https://pure.uva.nl/ws/files/42206706/Thesis.pdf}
\showURL{%
\tempurl}


\bibitem[Gundry(2015)]%
        {gundry2015}
\bibfield{author}{\bibinfo{person}{Adam Gundry}.}
  \bibinfo{year}{2015}\natexlab{}.
\newblock \showarticletitle{A Typechecker Plugin for Units of Measure:
  Domain-Specific Constraint Solving in {GHC} {Haskell}}. In
  \bibinfo{booktitle}{\emph{Proceedings of the 2015 {ACM} {SIGPLAN} Symposium
  on Haskell ({Haskell} '15)}}. \bibinfo{publisher}{ACM},
  \bibinfo{pages}{11--22}.
\newblock
\urldef\tempurl%
\url{https://doi.org/10.1145/2804302.2804305}
\showDOI{\tempurl}


\bibitem[Hart(1995)]%
        {hart1995}
\bibfield{author}{\bibinfo{person}{George~W. Hart}.}
  \bibinfo{year}{1995}\natexlab{}.
\newblock \bibinfo{booktitle}{\emph{Multidimensional Analysis: Algebras and
  Systems for Science and Engineering}}.
\newblock \bibinfo{publisher}{Springer}, \bibinfo{address}{New York}.
\newblock
\urldef\tempurl%
\url{https://doi.org/10.1007/978-1-4612-4208-6}
\showDOI{\tempurl}


\bibitem[Haynes(2026)]%
        {haynes2026}
\bibfield{author}{\bibinfo{person}{Houston Haynes}.}
  \bibinfo{year}{2026}\natexlab{}.
\newblock \bibinfo{title}{Dimensional Type Systems and Deterministic Memory
  Management: Design-Time Semantic Preservation in Native Compilation}.
\newblock \bibinfo{howpublished}{arXiv:2603.16437}.
\newblock
\urldef\tempurl%
\url{https://arxiv.org/abs/2603.16437v8}
\showURL{%
\tempurl}
\newblock
\shownote{Version 8, revised October 8, 2026}.


\bibitem[House(1983)]%
        {house1983}
\bibfield{author}{\bibinfo{person}{R.~T. House}.}
  \bibinfo{year}{1983}\natexlab{}.
\newblock \showarticletitle{A Proposal for an Extended Form of Type Checking of
  Expressions}.
\newblock \bibinfo{journal}{\emph{Comput. J.}} \bibinfo{volume}{26},
  \bibinfo{number}{4} (\bibinfo{year}{1983}), \bibinfo{pages}{366--374}.
\newblock
\urldef\tempurl%
\url{https://doi.org/10.1093/comjnl/26.4.366}
\showDOI{\tempurl}


\bibitem[{IEEE}(2019)]%
        {ieee754-2019}
\bibfield{author}{\bibinfo{person}{{IEEE}}.} \bibinfo{year}{2019}\natexlab{}.
\newblock \bibinfo{title}{{IEEE} Standard for Floating-Point Arithmetic}.
\newblock \bibinfo{howpublished}{IEEE Std 754-2019}.
\newblock
\urldef\tempurl%
\url{https://doi.org/10.1109/IEEESTD.2019.8766229}
\showDOI{\tempurl}


\bibitem[Jiang and Su(2006)]%
        {jiangsu2006}
\bibfield{author}{\bibinfo{person}{Lingxiao Jiang} {and}
  \bibinfo{person}{Zhendong Su}.} \bibinfo{year}{2006}\natexlab{}.
\newblock \showarticletitle{Osprey: A Practical Type System for Validating
  Dimensional Unit Correctness of {C} Programs}. In
  \bibinfo{booktitle}{\emph{Proceedings of the 28th International Conference on
  Software Engineering ({ICSE} '06)}}. \bibinfo{publisher}{ACM},
  \bibinfo{pages}{262--271}.
\newblock
\urldef\tempurl%
\url{https://doi.org/10.1145/1134285.1134323}
\showDOI{\tempurl}


\bibitem[Karr and {Loveman~III}(1978)]%
        {karrloveman1978}
\bibfield{author}{\bibinfo{person}{Michael Karr} {and}
  \bibinfo{person}{David~B. {Loveman~III}}.} \bibinfo{year}{1978}\natexlab{}.
\newblock \showarticletitle{Incorporation of Units into Programming Languages}.
\newblock \bibinfo{journal}{\emph{Commun. ACM}} \bibinfo{volume}{21},
  \bibinfo{number}{5} (\bibinfo{year}{1978}), \bibinfo{pages}{385--391}.
\newblock
\urldef\tempurl%
\url{https://doi.org/10.1145/359488.359501}
\showDOI{\tempurl}


\bibitem[Kennedy(1996)]%
        {kennedy1996}
\bibfield{author}{\bibinfo{person}{Andrew~J. Kennedy}.}
  \bibinfo{year}{1996}\natexlab{}.
\newblock \bibinfo{booktitle}{\emph{Programming Languages and Dimensions}}.
\newblock \bibinfo{type}{{T}echnical {R}eport} UCAM-CL-TR-391.
  \bibinfo{institution}{University of Cambridge, Computer Laboratory}.
\newblock
\urldef\tempurl%
\url{https://doi.org/10.48456/tr-391}
\showDOI{\tempurl}
\newblock
\shownote{PhD thesis, submitted November 1995}.


\bibitem[Kennedy(1997)]%
        {kennedy1997}
\bibfield{author}{\bibinfo{person}{Andrew~J. Kennedy}.}
  \bibinfo{year}{1997}\natexlab{}.
\newblock \showarticletitle{Relational Parametricity and Units of Measure}. In
  \bibinfo{booktitle}{\emph{Proceedings of the 24th {ACM} {SIGPLAN-SIGACT}
  Symposium on Principles of Programming Languages ({POPL} '97)}}.
  \bibinfo{publisher}{ACM}, \bibinfo{pages}{442--455}.
\newblock
\urldef\tempurl%
\url{https://doi.org/10.1145/263699.263761}
\showDOI{\tempurl}


\bibitem[Kennedy(2008)]%
        {kennedy2008}
\bibfield{author}{\bibinfo{person}{Andrew~J. Kennedy}.}
  \bibinfo{year}{2008}\natexlab{}.
\newblock \bibinfo{title}{Formalizing an Extensional Semantics for
  Units-of-Measure}.
\newblock \bibinfo{howpublished}{Talk, 3rd Informal {ACM} {SIGPLAN} Workshop on
  Mechanizing Metatheory ({WMM} '08)}.
\newblock
\urldef\tempurl%
\url{https://www.cis.upenn.edu/~sweirich/wmm/wmm08/kennedy.pdf}
\showURL{%
\tempurl}
\newblock
\shownote{Slides; they describe the accompanying {Coq} development}.


\bibitem[Kennedy(2010)]%
        {kennedy2009}
\bibfield{author}{\bibinfo{person}{Andrew~J. Kennedy}.}
  \bibinfo{year}{2010}\natexlab{}.
\newblock \showarticletitle{Types for Units-of-Measure: Theory and Practice}.
\newblock In \bibinfo{booktitle}{\emph{Central European Functional Programming
  School ({CEFP} 2009)}}. \bibinfo{series}{LNCS}, Vol.~\bibinfo{volume}{6299}.
  \bibinfo{publisher}{Springer}, \bibinfo{pages}{268--305}.
\newblock
\urldef\tempurl%
\url{https://doi.org/10.1007/978-3-642-17685-2_8}
\showDOI{\tempurl}


\bibitem[Lawson et~al\mbox{.}(1979)]%
        {lawson1979}
\bibfield{author}{\bibinfo{person}{Charles~L. Lawson},
  \bibinfo{person}{Richard~J. Hanson}, \bibinfo{person}{David~R. Kincaid},
  {and} \bibinfo{person}{Fred~T. Krogh}.} \bibinfo{year}{1979}\natexlab{}.
\newblock \showarticletitle{Basic Linear Algebra Subprograms for {Fortran}
  Usage}.
\newblock \bibinfo{journal}{\emph{ACM Trans. Math. Software}}
  \bibinfo{volume}{5}, \bibinfo{number}{3} (\bibinfo{year}{1979}),
  \bibinfo{pages}{308--323}.
\newblock
\urldef\tempurl%
\url{https://doi.org/10.1145/355841.355847}
\showDOI{\tempurl}


\bibitem[{Mars Climate Orbiter Mishap Investigation Board}(1999)]%
        {stephenson1999}
\bibfield{author}{\bibinfo{person}{{Mars Climate Orbiter Mishap Investigation
  Board}}.} \bibinfo{year}{1999}\natexlab{}.
\newblock \bibinfo{booktitle}{\emph{Mars Climate Orbiter Mishap Investigation
  Board: Phase {I} Report}}.
\newblock \bibinfo{type}{{T}echnical {R}eport}. \bibinfo{institution}{NASA}.
\newblock
\urldef\tempurl%
\url{https://llis.nasa.gov/llis_lib/pdf/1009464main1_0641-mr.pdf}
\showURL{%
\tempurl}
\newblock
\shownote{November 10, 1999; chaired by Arthur G. Stephenson}.


\bibitem[McBride and {Nordvall Forsberg}(2022)]%
        {mcbride2022}
\bibfield{author}{\bibinfo{person}{Conor McBride} {and}
  \bibinfo{person}{Fredrik {Nordvall Forsberg}}.}
  \bibinfo{year}{2022}\natexlab{}.
\newblock \showarticletitle{Type Systems for Programs Respecting Dimensions}.
\newblock In \bibinfo{booktitle}{\emph{Advanced Mathematical and Computational
  Tools in Metrology and Testing {XII}}},
  \bibfield{editor}{\bibinfo{person}{F.~Pavese}, \bibinfo{person}{A.~B.
  Forbes}, \bibinfo{person}{N.~F. Zhang}, {and} \bibinfo{person}{A.~G.
  Chunovkina}} (Eds.). \bibinfo{series}{Series on Advances in Mathematics for
  Applied Sciences}, Vol.~\bibinfo{volume}{90}. \bibinfo{publisher}{World
  Scientific}, \bibinfo{pages}{331--345}.
\newblock
\urldef\tempurl%
\url{https://doi.org/10.1142/9789811242380_0020}
\showDOI{\tempurl}


\bibitem[{National Institute of Standards and Technology} and {National Oceanic
  and Atmospheric Administration}(2020)]%
        {surveyfoot2020}
\bibfield{author}{\bibinfo{person}{{National Institute of Standards and
  Technology}} {and} \bibinfo{person}{{National Oceanic and Atmospheric
  Administration}}.} \bibinfo{year}{2020}\natexlab{}.
\newblock \showarticletitle{Deprecation of the United States ({U.S.}) Survey
  Foot}.
\newblock \bibinfo{journal}{\emph{Federal Register}} \bibinfo{volume}{85},
  \bibinfo{number}{193} (\bibinfo{year}{2020}), \bibinfo{pages}{62698--62708}.
\newblock
\newblock
\shownote{FR Doc. 2020-21902, October 5, 2020; deprecation effective December
  31, 2022}.


\bibitem[Orchard et~al\mbox{.}(2019)]%
        {orchard2019}
\bibfield{author}{\bibinfo{person}{Dominic Orchard},
  \bibinfo{person}{Vilem-Benjamin Liepelt}, {and} \bibinfo{person}{Harley
  {Eades III}}.} \bibinfo{year}{2019}\natexlab{}.
\newblock \showarticletitle{Quantitative Program Reasoning with Graded Modal
  Types}.
\newblock \bibinfo{journal}{\emph{Proceedings of the ACM on Programming
  Languages}} \bibinfo{volume}{3}, \bibinfo{number}{ICFP}
  (\bibinfo{year}{2019}), \bibinfo{pages}{110:1--110:30}.
\newblock
\urldef\tempurl%
\url{https://doi.org/10.1145/3341714}
\showDOI{\tempurl}


\bibitem[Ore et~al\mbox{.}(2017)]%
        {ore2017}
\bibfield{author}{\bibinfo{person}{John-Paul Ore}, \bibinfo{person}{Carrick
  Detweiler}, {and} \bibinfo{person}{Sebastian Elbaum}.}
  \bibinfo{year}{2017}\natexlab{}.
\newblock \showarticletitle{Lightweight Detection of Physical Unit
  Inconsistencies without Program Annotations}. In
  \bibinfo{booktitle}{\emph{Proceedings of the 26th {ACM} {SIGSOFT}
  International Symposium on Software Testing and Analysis ({ISSTA} '17)}}.
  \bibinfo{publisher}{ACM}, \bibinfo{pages}{341--351}.
\newblock
\urldef\tempurl%
\url{https://doi.org/10.1145/3092703.3092722}
\showDOI{\tempurl}


\bibitem[Owens et~al\mbox{.}(2016)]%
        {owens2016}
\bibfield{author}{\bibinfo{person}{Scott Owens}, \bibinfo{person}{Magnus~O.
  Myreen}, \bibinfo{person}{Ramana Kumar}, {and} \bibinfo{person}{Yong~Kiam
  Tan}.} \bibinfo{year}{2016}\natexlab{}.
\newblock \showarticletitle{Functional Big-Step Semantics}. In
  \bibinfo{booktitle}{\emph{Programming Languages and Systems ({ESOP} 2016)}}
  \emph{(\bibinfo{series}{LNCS}, Vol.~\bibinfo{volume}{9632})}.
  \bibinfo{publisher}{Springer}, \bibinfo{pages}{589--615}.
\newblock
\urldef\tempurl%
\url{https://doi.org/10.1007/978-3-662-49498-1_23}
\showDOI{\tempurl}


\bibitem[Paszke et~al\mbox{.}(2021)]%
        {paszke2021}
\bibfield{author}{\bibinfo{person}{Adam Paszke}, \bibinfo{person}{Daniel~D.
  Johnson}, \bibinfo{person}{David Duvenaud}, \bibinfo{person}{Dimitrios
  Vytiniotis}, \bibinfo{person}{Alexey Radul}, \bibinfo{person}{Matthew~J.
  Johnson}, \bibinfo{person}{Jonathan Ragan-Kelley}, {and}
  \bibinfo{person}{Dougal Maclaurin}.} \bibinfo{year}{2021}\natexlab{}.
\newblock \showarticletitle{Getting to the Point: Index Sets and
  Parallelism-Preserving Autodiff for Pointful Array Programming}.
\newblock \bibinfo{journal}{\emph{Proceedings of the ACM on Programming
  Languages}} \bibinfo{volume}{5}, \bibinfo{number}{ICFP}
  (\bibinfo{year}{2021}), \bibinfo{pages}{88:1--88:29}.
\newblock
\urldef\tempurl%
\url{https://doi.org/10.1145/3473593}
\showDOI{\tempurl}


\bibitem[Petricek et~al\mbox{.}(2014)]%
        {petricek2014}
\bibfield{author}{\bibinfo{person}{Tomas Petricek}, \bibinfo{person}{Dominic
  Orchard}, {and} \bibinfo{person}{Alan Mycroft}.}
  \bibinfo{year}{2014}\natexlab{}.
\newblock \showarticletitle{Coeffects: A Calculus of Context-Dependent
  Computation}. In \bibinfo{booktitle}{\emph{Proceedings of the 19th {ACM}
  {SIGPLAN} International Conference on Functional Programming ({ICFP} '14)}}.
  \bibinfo{publisher}{ACM}, \bibinfo{pages}{123--135}.
\newblock
\urldef\tempurl%
\url{https://doi.org/10.1145/2628136.2628160}
\showDOI{\tempurl}


\bibitem[Poulsen et~al\mbox{.}(2018)]%
        {poulsen2018}
\bibfield{author}{\bibinfo{person}{Casper~Bach Poulsen}, \bibinfo{person}{Arjen
  Rouvoet}, \bibinfo{person}{Andrew Tolmach}, \bibinfo{person}{Robbert
  Krebbers}, {and} \bibinfo{person}{Eelco Visser}.}
  \bibinfo{year}{2018}\natexlab{}.
\newblock \showarticletitle{Intrinsically-Typed Definitional Interpreters for
  Imperative Languages}.
\newblock \bibinfo{journal}{\emph{Proceedings of the ACM on Programming
  Languages}} \bibinfo{volume}{2}, \bibinfo{number}{POPL}
  (\bibinfo{year}{2018}), \bibinfo{pages}{16:1--16:34}.
\newblock
\urldef\tempurl%
\url{https://doi.org/10.1145/3158104}
\showDOI{\tempurl}


\bibitem[Reynolds(1972)]%
        {reynolds1972}
\bibfield{author}{\bibinfo{person}{John~C. Reynolds}.}
  \bibinfo{year}{1972}\natexlab{}.
\newblock \showarticletitle{Definitional Interpreters for Higher-Order
  Programming Languages}. In \bibinfo{booktitle}{\emph{Proceedings of the {ACM}
  Annual Conference, Volume 2}}. \bibinfo{publisher}{ACM},
  \bibinfo{pages}{717--740}.
\newblock
\urldef\tempurl%
\url{https://doi.org/10.1145/800194.805852}
\showDOI{\tempurl}


\bibitem[Russell(1903)]%
        {russell1903}
\bibfield{author}{\bibinfo{person}{Bertrand Russell}.}
  \bibinfo{year}{1903}\natexlab{}.
\newblock \bibinfo{booktitle}{\emph{The Principles of Mathematics}}.
\newblock \bibinfo{publisher}{Cambridge University Press},
  \bibinfo{address}{Cambridge}.
\newblock


\bibitem[Tait(1967)]%
        {tait1967}
\bibfield{author}{\bibinfo{person}{William~W. Tait}.}
  \bibinfo{year}{1967}\natexlab{}.
\newblock \showarticletitle{Intensional Interpretations of Functionals of
  Finite Type {I}}.
\newblock \bibinfo{journal}{\emph{Journal of Symbolic Logic}}
  \bibinfo{volume}{32}, \bibinfo{number}{2} (\bibinfo{year}{1967}),
  \bibinfo{pages}{198--212}.
\newblock
\urldef\tempurl%
\url{https://doi.org/10.2307/2271658}
\showDOI{\tempurl}


\bibitem[Tobin-Hochstadt and Allen(2005)]%
        {tobinhochstadt2005}
\bibfield{author}{\bibinfo{person}{Sam Tobin-Hochstadt} {and}
  \bibinfo{person}{Eric Allen}.} \bibinfo{year}{2005}\natexlab{}.
\newblock \showarticletitle{A Core Calculus of Metaclasses}. In
  \bibinfo{booktitle}{\emph{Foundations of Object-Oriented Languages ({FOOL}
  '05)}}.
\newblock
\urldef\tempurl%
\url{https://homepages.inf.ed.ac.uk/wadler/fool/program/final/7/7_Paper.pdf}
\showURL{%
\tempurl}


\bibitem[{Tranc{\'o}n y Widemann} and Lepper(2025)]%
        {tranconlepper2025}
\bibfield{author}{\bibinfo{person}{Baltasar {Tranc{\'o}n y Widemann}} {and}
  \bibinfo{person}{Markus Lepper}.} \bibinfo{year}{2025}\natexlab{}.
\newblock \showarticletitle{A Theory of Conversion Relations for Prefixed Units
  of Measure}.
\newblock \bibinfo{journal}{\emph{Fundamenta Informaticae}}
  \bibinfo{volume}{195}, \bibinfo{number}{1--4}, Article
  \bibinfo{articleno}{11} (\bibinfo{year}{2025}), \bibinfo{numpages}{45}~pages.
\newblock
\urldef\tempurl%
\url{https://doi.org/10.46298/fi.12436}
\showDOI{\tempurl}


\bibitem[Wadler and Blott(1989)]%
        {wadlerblott1989}
\bibfield{author}{\bibinfo{person}{Philip Wadler} {and}
  \bibinfo{person}{Stephen Blott}.} \bibinfo{year}{1989}\natexlab{}.
\newblock \showarticletitle{How to Make Ad-Hoc Polymorphism Less Ad Hoc}. In
  \bibinfo{booktitle}{\emph{Proceedings of the 16th {ACM} {SIGPLAN-SIGACT}
  Symposium on Principles of Programming Languages ({POPL} '89)}}.
  \bibinfo{publisher}{ACM}, \bibinfo{pages}{60--76}.
\newblock
\urldef\tempurl%
\url{https://doi.org/10.1145/75277.75283}
\showDOI{\tempurl}


\bibitem[Wand and O'Keefe(1991)]%
        {wandokeefe1991}
\bibfield{author}{\bibinfo{person}{Mitchell Wand} {and}
  \bibinfo{person}{Patrick~M. O'Keefe}.} \bibinfo{year}{1991}\natexlab{}.
\newblock \showarticletitle{Automatic Dimensional Inference}.
\newblock In \bibinfo{booktitle}{\emph{Computational Logic: Essays in Honor of
  Alan Robinson}}, \bibfield{editor}{\bibinfo{person}{Jean-Louis Lassez} {and}
  \bibinfo{person}{Gordon Plotkin}} (Eds.). \bibinfo{publisher}{MIT Press},
  \bibinfo{pages}{479--486}.
\newblock


\end{thebibliography}

\end{document}